\documentclass[twocolumn,showpacs,amsmath,amssymb]{revtex4}
\usepackage[dvips]{graphics}
\def\imo{i}
\def\re#1{Re(#1)}
\def\im#1{Im(#1)}

\begin{document}
\title{A massive charged scalar field in the Kerr-Newman background I: quasinormal modes, late-time tails and stability}
\author{R. A. Konoplya}\email{konoplya_roma@yahoo.com}
\affiliation{DAMTP, Centre for Mathematical Sciences, University of Cambridge, Wilberforce Road, Cambridge CB3 0WA, UK.}
\author{A. Zhidenko}\email{olexandr.zhydenko@ufabc.edu.br}
\affiliation{Centro de Matem\'atica, Computa\c{c}\~ao e Cogni\c{c}\~ao, Universidade Federal do ABC (UFABC),\\ Rua Aboli\c{c}\~ao, CEP: 09210-180, Santo Andr\'e, SP, Brazil}

\begin{abstract}
  So far analysis of the quasinormal spectrum of a massive charged scalar field in the black hole background has been limited by the regime of small $\mu M$ and $qQ$, where $\mu$, $q$ ($M$, $Q$) are mass and charge of the field (black hole). Here we shall present a comprehensive picture of quasinormal modes, late-time tails and stability of a massive charged scalar field around Kerr-Newman black holes for any physically meaningful values of the parameters. We shall show that despite presence of the two mechanisms of superradiance (owing to black hole's rotation and charge) and the massive term creating growing bound states, there is no indication of instability under quasinormal modes' boundary conditions. We have shown that for some moderate values of $qQ$ dominant quasinormal modes may have \emph{arbitrarily small real oscillation frequencies} $Re(\omega)$. An analytic formula for the quasinormal modes has been derived in the regime of large $qQ$. The larger the field's charge, the sooner asymptotic tails dominate in a signal, making it difficult to extract quasinormal frequencies from a time-domain profile. Analytic expressions for intermediate and asymptotically late-time tails have been found for the Reissner-Nordstr\"om black hole. For the near extremal Kerr-Newman black holes we have obtained a more general picture of the \emph{mode branching} found recently for massless fields (arXiv:1212.3271) in the Kerr background.
\end{abstract}
\maketitle

\section{Introduction}

The general solution for an asymptotically flat black hole in the Einstein-Maxwell theory is given by the Kerr-Newman solution.
It describes an electrically charged rotating black hole with its three parameters: mass, charge and angular momentum.
A systematic study of the fields' dynamic in the vicinity of a black hole is essential for understanding black-hole evaporation, quasinormal modes and stability.
The linear dynamics of a charged massive scalar field in the background of a charged black hole is characterized by the two dimensionless parameters $\mu M$ and $qQ$.
A black hole is not believed to be formed possessing considerable electric charge, and, once it is formed, it undergoes a rather quick discharging \cite{Carter,Gibbons:1975kk}.
Yet, even if a black hole has a very small charge of order $10^2 e$, the parameter $qQ$ need not be small. In addition, a charge induced by an external magnetic field, may be formed at the surface of an initially neutral, but rotating black hole \cite{Gibbons-magn,Aliev:1989wx}. Thus, the complete analysis of a massive charged scalar field dynamics should include consideration of the whole range of values $q Q$ and $\mu M$.

In this work we shall study the stability and evolution of perturbation of a massive charged scalar field in the Kerr-Newman background in terms of its \emph{quasinormal modes} and \emph{asymptotic tails} at late times. It is believed that if the quasinormal modes are damped, the system under consideration is stable, though a rigorous mathematical proof of stability is usually complicated and sometimes includes a nontrivial analysis of the initial value problem. By now, quite a few papers have been devoted to scalar field perturbations in the black-hole background, yet, while the behavior of the massless neutral scalar field is indeed very well studied, the quasinormal modes of charged fields was studied only in the regime $\mu M \ll 1$, $q Q \ll 1$ \cite{Hod:1997mt,Konoplya:2002ky,Konoplya:2002wt,Jing:2007zza,Konoplya:2007zx,Kokkotas:2010zd,Flachi:2012nv}, except for \cite{Hod:2012zzb}, where the WKB estimation for quasinormal modes (QNMs) of a massless charged scalar field around the Reissner-Nordstr\"om black hole was given in the regime $\ell \ll qQ \ll \ell^2$, where $\ell$ is the multipole number. The complete analysis of quasinormal modes (allowing also to judge about stability) for a massive charged scalar field \emph{for arbitrary values} $q Q$ and $\mu M$ has been lacking so far not only for the Kerr-Newman, but even for Reissner-Nordstr\"om solutions. Here we shall solve this problem by adopting the two numerical methods of computation (the Frobenius method and the time-domain integrations) based on convergent procedures, which allow us to find quasinormal modes accurately and with no restriction on the parameters of the system.

Perturbation of a charged massive field in the background of a rotating charged black hole has rich physics, because there are a number of phenomena which must be taken into consideration:
\begin{itemize}
  \item \emph{Superradiance}, that is the amplification of waves with particular frequencies reflected by a black hole, if it is rotating \cite{Starobinsky} or electrically charged \cite{Bekenstein}. Thus, there will be the two regimes of superradiance for Kerr-Newman black holes, owing to charge and rotation \cite{Furuhashi:2004jk}.
  \item \emph{Superradiant instability} of bound states around black holes owing to the massive term, which creates a local minimum far from the black hole, so that the wave will be reflected repeatedly and can grow. It is essential that this instability occurs under the bound states' boundary condition, which differ from the quasinormal modes' ones \cite{Furuhashi:2004jk}.
  \item \emph{Quasiresonances.} When a field is massive, quasinormal modes with arbitrarily long lifetimes, called quasiresonances, appear, once some critical value of mass of the field is achieved \cite{quasi-resonances}. When the damping rate goes to zero, the quasinormal  asymptotically approach the bound state, but still remain quasinormal modes for whatever small but nonzero damping. \cite{quasi-resonances}.
  \item \emph{Instability of the extremal black holes} that apparently occurs for fields of any spin and both for extremal Reissner-Nordstr\"om and Kerr black holes \cite{Aretakis:2011ha,Aretakis:2011gz,Lucietti:2012xr,Lucietti:2012sf}, and therefore, must be expected for the extremal Kerr-Newman solution as well.  However, in the linear approximation this instability develops only on the event horizon and cannot be seen by an external observer.
  \item \emph{Mode branching}. Quasinormal modes of Kerr black holes were believed to be completely studied until a few months ago when an interesting observation has been made \cite{Yang:2012pj}. It was shown that, for the near-extremal rotation there are two distinct sets of damped quasinormal modes, which merge to a single set in the exactly extremal state \cite{Yang:2012pj}.
\end{itemize}

Here, through the numerical analysis of quasinormal modes and asymptotic tails we have shown that a massive charged scalar field is stable in the vicinity of the Kerr-Newman black hole, in spite of the instability of the corresponding bound states. We found that at some values of the field's charge $q$ quasinormal modes may behave qualitatively differently from those of the neutral field: the fundamental mode (dominating at late times) may have arbitrarily small real part (real oscillation frequency) which appears in the time domain as a very short period of quasinormal ringing consisting of damped oscillations and the quick onset of asymptotic power-law tails. In addition, we generalized earlier results on mode branching of massless fields around nearly extremal Kerr black holes to the massive fields and Kerr-Newman solutions. An analytic formula has been obtained for large $qQ$.

The paper is organized as follows.  In Sec II the basic formulas for a charged massive scalar field in the Kerr-Newman background is given. The wave equation is reduced to the Schr\"odinger-like form with an effective potential. Sec III describes the numerical methods which we used: the Frobenius methods, two schemes of time-domain integration (for neutral and charged fields) together with the method for extraction of frequencies from the time-domain profiles, called the Prony method, and the WKB approach. We have related separately perturbations of nonextremal black holes (Sec IV) and nearly and exactly extremal ones (Sec. V), as near extremal black holes shows new phenomena, such as mode branching.  In Sec. VI we discuss some technical difficulties which appear when one considers higher overtones of a charged scalar field or approach closely to the extremal state, keeping $q$ non zero. In Sec VII we summarize the results obtained.

\section{Kerr-Newman background}

In the Boyer-Lindquist coordinates the Kerr-Newman metric has the form
\begin{eqnarray}
ds^2 &=& \frac{\Delta_r}{ \rho^2}(dt-a\sin^2\theta d\varphi)^2-\rho^2
\left(\frac{dr^2}{\Delta_r}+\frac{d\theta^2}{\Delta_\theta}\right)\\\nonumber
&&-\frac{\Delta_\theta \sin^2\theta}{\rho^2}
[adt-(r^2+a^2)d\varphi]^2,
\end{eqnarray}
where
$$\Delta_r=(r^2+a^2)-2Mr+Q^2,$$
\begin{equation}
\rho^2=r^2+a^2\cos^2\theta,
\end{equation}
and $Q$ is the black-hole charge, $M$ is its mass. The electromagnetic background of the black hole is given by the four-vector potential,
\begin{equation}
A_{\mu}dx^{\mu}
   =-\frac{Qr}{\rho^2}(dt-a\sin^2\theta d\varphi).
\end{equation}
We shall parameterize the metric by the following three parameters: the event horizon $r_+$, the inner horizon $r_-$, and the rotation parameter $a$,
$$0\leq a^2/r_+\leq r_-\leq r_+.$$
The black hole's mass and charge are then
$$2M=r_++r_-,\qquad Q^2=r_+r_--a^2.$$

A massive charged scalar field satisfies the Klein-Gordon equation,
\begin{eqnarray}\label{KG}
&&\frac{1}{\sqrt{-g}}\frac{\partial}{\partial x^\alpha}\left(g^{\alpha\beta}\sqrt{-g}\frac{\partial\psi}{\partial x^\beta}\right)+2iqA_\alpha g^{\alpha\beta}\frac{\partial\psi}{\partial x^\beta}\\\nonumber&&+(\mu^2-q^2g^{\alpha\beta}A_\alpha A_\beta)\psi=0,
\end{eqnarray}
where $q$ and $\mu$ are the field's charge and mass respectively.

One separates variables by the following ansatz
\begin{equation}\label{anzats}
\psi = e^{-\imo \omega t + \imo m \phi} S(\theta) R(r)/\sqrt{r^2+a^2},
\end{equation}
where $S(\theta)$ obeys the following equation
\begin{eqnarray}\label{angularpart}
&&\left(\frac{\partial^2}{\partial \theta^2} + \cot \theta \frac{\partial}{\partial\theta} - \frac{m^2}{\sin^2 \theta} - a^2 \omega^2\sin^2 \theta\right.
\\\nonumber&&\left. + 2 m a \omega + \lambda - \mu^2 a^2 \cos^2 \theta \right) S(\theta) = 0,
\end{eqnarray}
and $\lambda$ is the separation constant.

This equation can be solved numerically for any value of $\omega$ in the same way as the equation for a massive scalar field in the Kerr black-hole background \cite{Konoplya:2006br}. Let us note that, when $\mu=0$, Eq.~(\ref{angularpart}) can be reduced to the well-known equation for the spheroidal functions. In this case, for any fixed value of $\omega$ the separation constant $\lambda$ can be found numerically using the continued fraction method \cite{Suzuki:1998vy}. When the effective mass is not zero, the separation constant $\lambda(\omega, \mu)$ can be expressed, in terms of the eigenvalue for spheroidal functions $\lambda(\omega)$ \cite{Kokkotas:2010zd}, as
$$\lambda(\omega,\mu)=\lambda(\sqrt{\omega^2-\mu^2},0)+2ma(\sqrt{\omega^2-\mu^2}-\omega)+\mu^2a^2.$$
When $a=0$, one has $\lambda=\ell(\ell+1),~\ell=0,1,2\ldots$. For nonzero values of $a$, the separation constant can be enumerated by the integer multipole number $\ell\geq|m|$.

The radial function satisfies a Schr\"odinger equation,
\begin{equation}\label{radialpart}
\left(\frac{d^2}{d x^2} - V(x)\right) R(x) = 0,
\end{equation}
where $x$ is the tortoise coordinate,
$$dx=\frac{(r^2+a^2)}{\Delta_r}dr,$$
and the effective potential is
\begin{eqnarray}\nonumber
V&=&\frac{\Delta_r}{(r^2+a^2)^2}\left(\lambda + \mu^2 r^2+\frac{(r \Delta_r)'}{r^2 + a^2}-\frac{3 \Delta_r r^2}{(r^2 + a^2)^2}\right)
\\\label{Effective-potential}&&-\left(\omega - \frac{m a+qQr}{r^2 + a^2}\right)^2.
\end{eqnarray}

The asymptotics of the effective potential near the event horizon and at spatial infinity are
\begin{equation}
\begin{array}{lll}
V \rightarrow -\Omega^2, &~~r \rightarrow \infty~~(x\rightarrow \infty),&  \Omega = \sqrt{\omega^2 - \mu^2},\\
V \rightarrow -\tilde\omega^2, &~~r \rightarrow r_+~~(x\rightarrow -\infty),& \displaystyle \tilde\omega = \omega - \frac{ma+qQr_+}{a^2+r_+^2},
\end{array}
\end{equation}
where we fix the sign of $\Omega$ such that $\re{\Omega}$ is of the same sign as $\re{\omega}$.

Note, that $\re{\tilde\omega}$ and $\re{\omega}$ can have different signs. This corresponds to the superradiant regime in which 
one has
$$\begin{array}{lll}
      0<\re{\omega} < \displaystyle\frac{ma+qQr_+}{a^2+r_+^2} & \hbox{or} & \displaystyle\frac{ma+qQr_+}{a^2+r_+^2}<\re{\omega} < 0.
    \end{array}$$

By definition, quasinormal modes (QNMs) are proper oscillation frequencies which correspond to purely incoming wave at the event horizon and purely outgoing wave at infinity, so that no incoming waves from either of the ``infinities'' are allowed.
Thus, the boundary conditions for the QNMs can be written as follows
\begin{equation}\label{boundary-conditions}
\begin{array}{ll}
R \propto \exp(\imo\Omega x),&\quad x\rightarrow \infty,\\
R \propto \exp(-\imo\tilde\omega x),&\quad x\rightarrow -\infty.
\end{array}
\end{equation}

\section{Numerical techniques}

Here, we shall briefly relate the three numerical methods used for finding quasinormal frequencies:
\begin{itemize}
\item Leaver method, which is based on a convergent procedure and, thereby, allowing one to find QN modes accurately,
\item WKB method (accurate in the regime of high multipole numbers),
\item time-domain integration which includes contribution of all modes, and, together with the Prony method, usually allows extracting a few lower dominant frequencies from a time-domain profile.
\end{itemize}
We shall see that in some ranges of parameters, when one method becomes slowly convergent or inapplicable, the other can be used, so that the use of a few alternative methods is necessary here not only for an additional checking, but also for getting the complete picture of quasinormal modes and stability in the full range of parameters.

\subsection{Leaver method}
Equation (\ref{radialpart}) has an irregular singularity at spatial infinity and four regular singularities at $r=r_+$, $r=r_-=(Q^2+a^2)/r_+$ and $r=\pm\imo a$. The appropriate Frobenius series is determined as
$$R(r)=\left(\frac{r-r_+}{r-r_-}\right)^{-\displaystyle\imo\tilde\omega/4\pi T_H}e^{\displaystyle\imo\Omega r}(r-r_-)^{\displaystyle\imo\sigma}y(r),$$
where
$$\sigma=\left(\Omega+\frac{\mu^2}{2\Omega}\right)(r_++r_-),$$
and $T_H$ is the Hawking temperature
$$T_H=\frac{\Delta'(r_+)}{4\pi(r_+^2+a^2)}.$$
The function $y(r)$ must be regular at the horizon and spatial infinity, so that the series in the vicinity of the event horizon
$$y(r)=\frac{\sqrt{r^2+a^2}}{r-r_-}\sum_{k=0}^{\infty}a_k(\omega)\left(\frac{r-r_+}{r-r_-}\right)^k$$
satisfies both these requirements.
This series converges everywhere outside the event horizon ($r_+\leq r<\infty$). When boundary conditions (\ref{boundary-conditions}) are satisfied, that is when $\omega$ is a quasinormal (QN) frequency, the series convergence also at the spatial infinity \cite{Leaver:1985ax}.
The coefficients $a_k$ satisfy the three-term recurrence relation.
\begin{equation}
\alpha_n a_{n+1} + \beta_n a_n + \gamma_n a_{n-1} = 0, \quad n \geq 0, \qquad \gamma_0 = 0,
\end{equation}
where $\alpha_n$, $\beta_n$, $\gamma_n$ can be found in an analytic form.

By comparing the ratio of the series coefficients
\begin{eqnarray}%
\frac{a_{n+1}}{a_n}&=&\frac{\gamma_{n}}{\alpha_n}\frac{\alpha_{n-1}}{\beta_{n-1}
-\frac{\alpha_{n-2}\gamma_{n-1}}{\beta_{n-2}-\alpha_{n-3}\gamma_{n-2}/\ldots}}-\frac{\beta_n}{\alpha_n},\nonumber\\
\label{ratio}\frac{a_{n+1}}{a_n}&=&-\frac{\gamma_{n+1}}{\beta_{n+1}-\frac{\alpha_{n+1}\gamma_{n+2}}{\beta_{n+2}-\alpha_{n+2}\gamma_{n+3}/\ldots}},
\end{eqnarray}%
we obtain an equation with a convergent \emph{infinite continued fraction} on its right side:
\begin{eqnarray}\label{continued_fraction} \beta_n-\frac{\alpha_{n-1}\gamma_{n}}{\beta_{n-1}
-\frac{\alpha_{n-2}\gamma_{n-1}}{\beta_{n-2}-\alpha_{n-3}\gamma_{n-2}/\ldots}}=\qquad\\\nonumber
\frac{\alpha_n\gamma_{n+1}}{\beta_{n+1}-\frac{\alpha_{n+1}\gamma_{n+2}}{\beta_{n+2}-\alpha_{n+2}\gamma_{n+3}/\ldots}},
\end{eqnarray}%
which can be solved numerically by minimizing the absolute value of the difference between its left- and right-hand sides. Equation (\ref{continued_fraction}) has an infinite number of roots, but the most stable root depends on $n$. The larger number $n$ corresponds to the larger imaginary part of the root $\omega$ \cite{Leaver:1985ax}. As we study QNMs with slower decay rate, we usually choose $n=0$.
In order to improve convergence of the infinite continued fraction for nonzero mass of a scalar field, we use the Nollert procedure \cite{Nollert}.

\subsection{WKB formula}
As for the study of mode branching we are also interested in modes with high multipole numbers, the WKB approach is useful here, which is accurate in the eikonal regime $\ell \gg n$ and usually provides very good accuracy at moderate $\ell > n$. The WKB formula for calculation of QNMs has the following form:
\begin{equation}\label{WKBformula}
	\frac{\imo V_{0}}{\sqrt{2 V_{0}''}} - \sum_{i=2}^{i=6}
		\Lambda_{i} = n+\frac{1}{2},\qquad n=0,1,2\ldots,
\end{equation}
where $V_{0}$ and $V_{0}''$ are the values of the effective potential (\ref{Effective-potential}) and its second derivative with respect to the tortoise coordinate $x$ at the potential's peak.  The terms $\Lambda_{i}$ depend on higher derivatives of $V$ at its maximum, and $n$ labels the overtones. The WKB approach was developed by Schutz and Will \cite{WKB} and later extended to higher orders \cite{WKBorder}. Since there is implicit dependence on $\omega$ (either through the separation constant $\lambda$ for nonvanishing rotation or due to the $qQ$ coupling in the effective potential (\ref{Effective-potential})), one has to search for the roots of equation (\ref{WKBformula}) by minimizing the absolute value of the difference between its left and right sides.

\subsection{Time-domain integration}

For nonrotating black holes we are able to construct a time-dependent profile of the wave function at a fixed $x$.
The recently found instability of the exactly extremal Reissner-Nordstr\"om black hole \cite{Aretakis:2011ha} makes it important to check stability of the nearly and exactly extremal black holes. Even when the Frobenius method finds no growing quasinormal modes, one could think that the growing modes were simply missed in the frequency domain at the stage of search for the roots of the equation with continued fractions Eq.~(\ref{continued_fraction}), or, that a different boundary condition should be imposed at the event horizon in the extremal case. In order to eliminate both these suspicions, we will use the numerical characteristic integration method in time domain. Here, two different schemes of integration were used for \emph{neutral} and \emph{charged} fields.

\textbf{A scheme for a neutral field.} The wave equation can be written in time-dependent form as follows:
\begin{equation}\label{wavelike}
\frac{\partial^2\Psi}{\partial t^2}-\frac{\partial^2\Psi}{\partial x^2}+V(t,x)\Psi=0.
\end{equation}
The technique of integration of the above wave equation in the time domain was developed in \cite{Gundlach:1993tp}. The method uses the light-cone variables $u = t - r^{*}$, $v = t + r^{*}$, so that the wave equation reads
\begin{equation}\label{light-cone}
\left(4\frac{\partial^2}{\partial u\partial v}+V(u,v)\right)\Psi(u,v)=0.
\end{equation}
The initial data are specified on the two null surfaces $u = u_0$ and $v = v_0$.
Acting by the time evolution operator $\exp\left(h\frac{\partial}{\partial t}\right)$  on $\Psi$ and taking account of (\ref{light-cone}), one finds
\begin{eqnarray}
\Psi(N)&=& \Psi(W)+\Psi(E)-\Psi(S)\nonumber\\
&& -\frac{h^2}{8}V(S)\left(\Psi(W)+\Psi(E)\right) + \mathcal{O}(h^4),\label{integration-scheme}
\end{eqnarray}
where one introduced letters to mark the points as follows: $S=(u,v)$, $W=(u+h,v)$, $E=(u,v+h)$, and $N=(u+h,v+h)$.
Equation (\ref{integration-scheme}) allows us to calculate the values of $\Psi$ inside the rhombus, which is built on the two null-surfaces $u=u_0$ and $v=v_0$, starting from the initial data specified on them. As a result we can find the time profile data $\{\Psi(t=t_0),\Psi(t=t_0+h),\Psi(t=t_0+2h),\ldots\}$ in each point of the rhombus.

The time-domain integration includes a contribution from all overtones, and, thus, missing some mode is excluded. This method is based on the scattering of the Gaussian wave on the potential barrier and therefore does not specify the boundary condition on the event horizon, so that the potentially missed instability due to possibly different boundary conditions must be also discarded.

\begin{figure*}
\resizebox{\linewidth}{!}{\includegraphics*{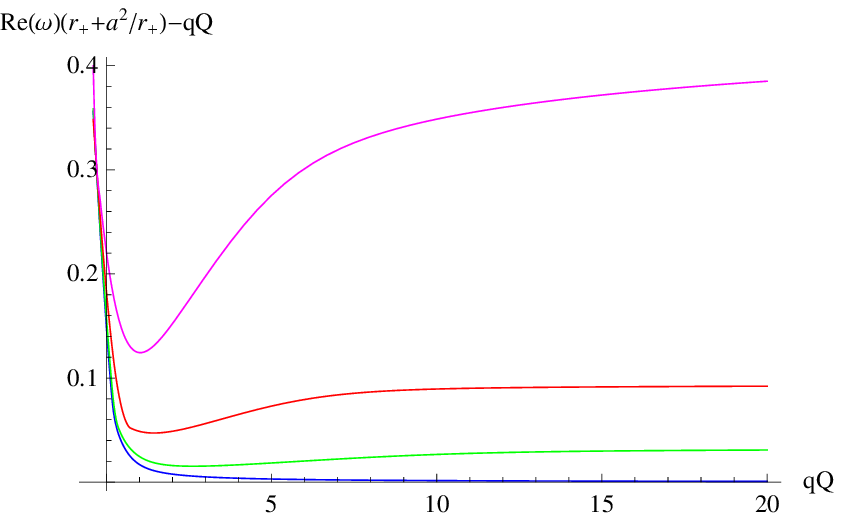}\includegraphics*{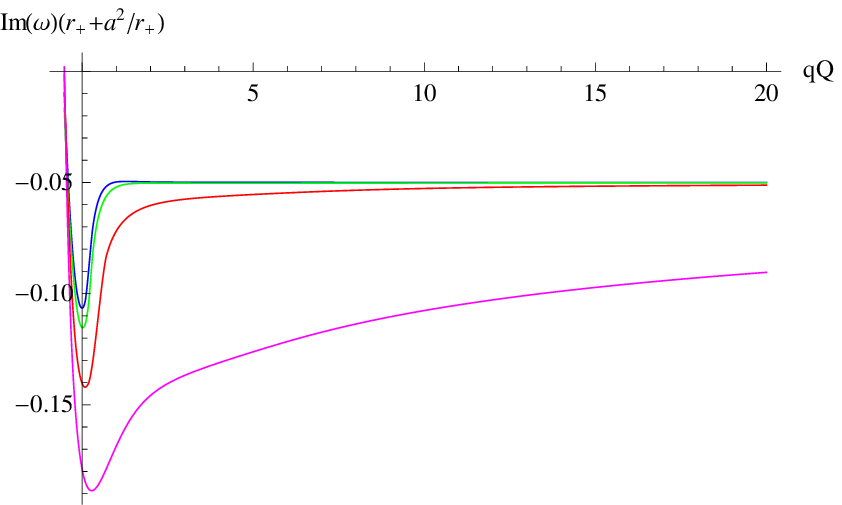}}
\caption{Real (left panel) and imaginary (right panel) parts of the fundamental ($\ell=n=0$) quasinormal frequency (in units of $r_++a^2/r_+$) as functions of $qQ$ for the Kerr-Newman black holes 
$r_-=0.8r_+$
with $a=0$ (blue), $a=0.3r_+$ (green), $a=0.6r_+$ (red), and $a^2=r_-r_+$ (magenta). The bigger rotation parameter ($a$) corresponds to the larger absolute values of the real and imaginary parts, and the asymptotical regime is achieved at a larger charge $q$.}\label{figasymptot}
\end{figure*}

\textbf{A scheme for a charged field.} When the scalar field is charged, an extra term containing the first derivative in time appears. Therefore, a different integration scheme is required. We shall use the finite difference scheme proposed in \cite{Abdalla:2010nq}. First we rewrite the wave-like equation (\ref{radialpart}) for the Reissner-Nordstr\"om black hole ($a=0$) in the time-dependent form
\begin{equation}
\left(\frac{\partial^2}{\partial x^2}-\frac{\partial^2}{\partial t^2}+2\imo\Phi(x)\frac{\partial}{\partial t} - V(x)\right)\Psi(t,x) = 0,
\end{equation}
where
\begin{eqnarray}\nonumber
V(r)&=&\frac{\Delta_r}{r^4}\left(\lambda + \mu^2 r^2+\frac{r\Delta_r'-2\Delta_r}{r^2}\right)-\Phi(r)^2\\
\Phi(r)&=&-\frac{qQ}{r}, \qquad \lambda=\ell(\ell+1)=0,2,6,24\ldots
\end{eqnarray}
Following \cite{Abdalla:2010nq}, one can derive the evolution of $\Psi$ in an isosceles triangle with the base on the axis $x$, where initial conditions are imposed. Then, one has
\begin{eqnarray}\Psi_{j,i+1}&=&\frac{(1+\imo\Phi_j\Delta t)\Psi_{j,i-1}+(2-\Delta t^2V_j)\Psi_{j,i}}{1-\imo\Phi_j\Delta t}\nonumber\\
&&+\frac{\Delta t^2}{\Delta x^2}\frac{\Psi_{j+1,i}+\Psi_{j-1,i}-2\Psi_{j,i}}{1-\imo\Phi_j\Delta t}\,.
\end{eqnarray}
Indexes $i$ and $j$ enumerate, respectively, the coordinates $t$ and $x$ of the grid:
$$x_i=x_0+i\Delta x,\qquad t_j=t_0+j\Delta t.$$
We choose the initial conditions again as a Gaussian distribution whose maximum is near the maximum of the effective potential. Since Von Neumann stability conditions require $\Delta t<\Delta x$, in this scheme we chose $\Delta x = 2\Delta t$. In order to achieve convergence we decrease $\Delta x$. Note, that as $qQ$ grows the convergence becomes seemingly slower, requiring smaller $\Delta x$, what increases the computation time.
When $q=0$ this scheme is reduced to the one above for the neutral field, yet, the codes for both schemes differ, so that letting $q=0$ in the Mathematica\textregistered{} code for the second scheme leads to a much longer computing than the first scheme.

\textbf{Prony method for mode extraction.} Once a time domain profile is found, one can extract dominant frequencies from it with the help of the Prony method. We fit the profile data by superposition of damped exponents
\begin{equation}\label{damped-exponents}
\Phi(t)\simeq\sum_{i=1}^pC_ie^{-\imo\omega_i t}
\end{equation}
and look for the convergence of the obtained frequencies at the increasing $p$. We shall show that although the fit works well for the neutral scalar field, it cannot be effectively used once modes with very small $\re{\omega}$ dominate in the spectrum at late times.

\section{Quasinormal modes and late-time tails of nonextremal black holes}

\subsection{Regime of large $qQ$}


In the regime of large $qQ$ and nonextremal $Q$, the Frobenius method allows us to find an approximate analytic expression for the quasinormal frequencies. When $qQ\gg1$ one can observe that $\alpha_n={\cal O}(qQ)$, $\beta_n={\cal O}(qQ)^2$, and $\gamma_n={\cal O}(qQ)^2$. Then, we can rewrite Eq.~(\ref{continued_fraction}) as
\begin{equation}\label{continued_fraction_qQ}
\frac{\beta_n}{(qQ)^2} + {\cal O}\left(\frac{1}{qQ}\right)=0.
\end{equation}
Considering $\lambda\ll (qQ)^2$, from (\ref{continued_fraction_qQ}) we find that $\omega={\cal O}(qQ)$. We observe that $\lambda\propto a\omega$ when $a\omega\gg1$. Then, we can write down $\lambda$ as $$\lambda=\lambda_0 a\omega+{\cal O}(1)$$ and find the asymptotic formula for the QNMs
\begin{eqnarray}\label{asymptotical_frequency}
\omega(a^2+r_+^2)=qQr_++a\left(m+\frac{r_+(r_+-r_-)}{r_+^2-a^2}\frac{\lambda_0}{4}\right)\\\nonumber-\imo(r_+-r_-)\frac{2n+1}{4}+{\cal O}\left(\frac{1}{qQ}\right)\,.
\end{eqnarray}
Here $n = 0, 1, 2, 3,\ldots$ is the overtone number. From the data given in Fig.~\ref{figasymptot} one can find that $\lambda_0\approx2$ for $\ell=0$.
In order to find $\lambda_0$ as a function of $\ell$ and $m$ we have analyzed numerically the asymptotic behavior of the eigenvalue of (\ref{angularpart}) and  found that
$$\lambda_0\approx4\left[\frac{\ell-m}{2}\right]+2,$$
where the brackets denote the integer part.

When $Q$ is not very close to its extremal value, the above analytical formula (in the limit of vanishing rotation) can be verified by the time-domain integration for a few lower modes, as it is shown in Fig.
\ref{figworotation}. There one can see that the analytic formula (\ref{asymptotical_frequency}) works very well already at moderate values of $qQ$. Nevertheless, even though we are able to obtain a stable time-domain profile for any values of $q$ and $Q$, the Prony method does not converge for high overtones as well as near extremal black holes. We shall discuss this in detail in the last section.
For $a=0$ (\ref{asymptotical_frequency}) coincides with the asymptotic formula found in \cite{Hod:2012zzb} in the regime $\ell\ll qQ\ll\ell^2$ with the help of the WKB approximation. Unlike \cite{Hod:2012zzb} in our calculations one does not need to be limited by  the regime $qQ\ll\ell^2$.

\subsection{Regime of small and moderate $q Q$}

\begin{figure*}
\resizebox{\linewidth}{!}{\includegraphics*{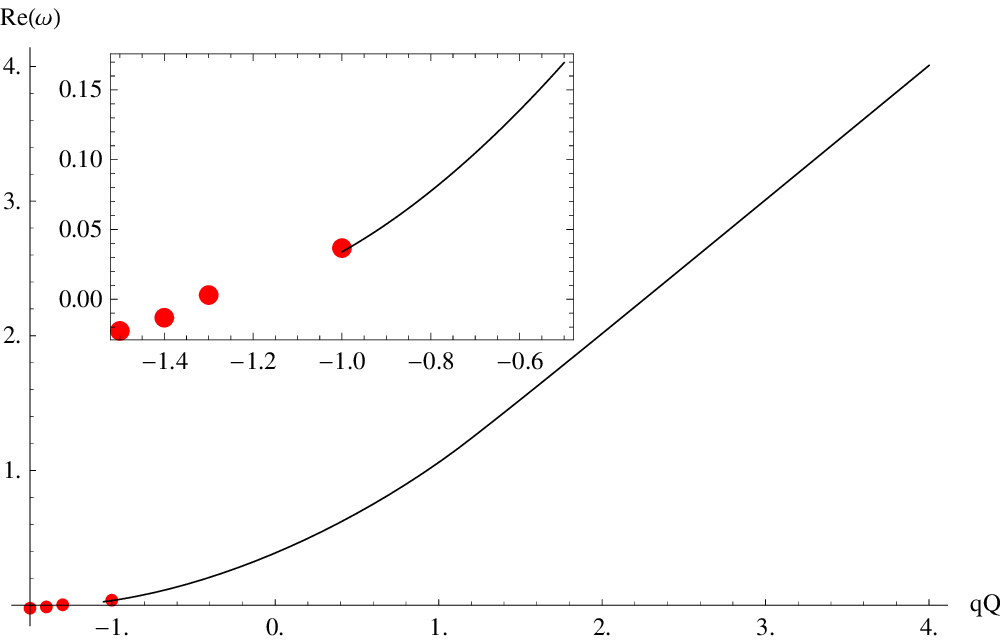}\includegraphics*{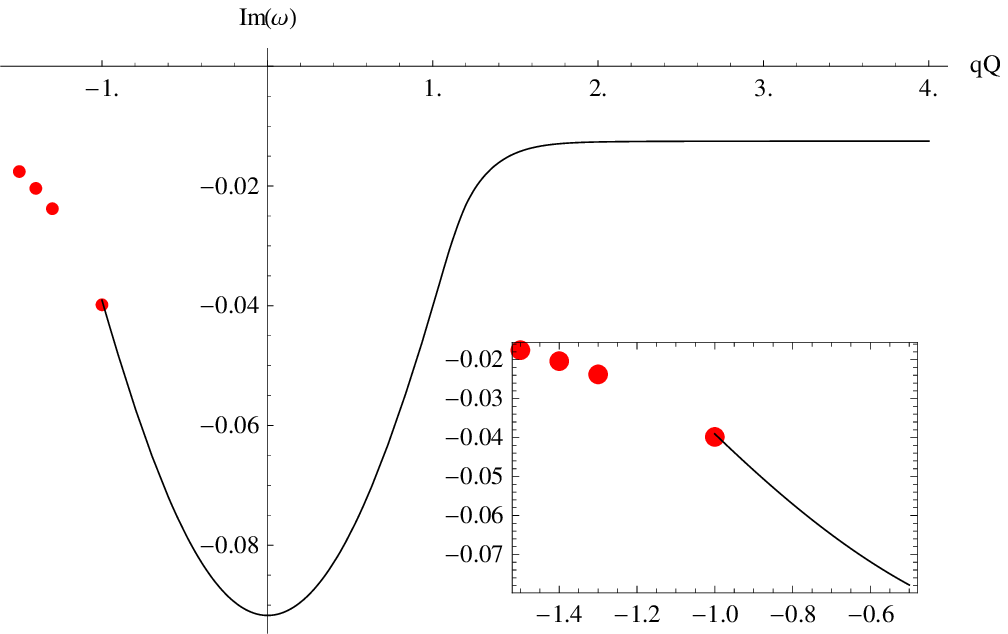}}
\caption{Real and imaginary part of the fundamental mode ($\ell=1$) of the massless scalar field as a function of $qQ$ for the quasi-extremal Reissner-Nordstr\"om black hole ($r_-=0.95$). Red dots correspond to the fit of time-domain profiles. Due to the symmetry of the equation the negative values of charge $q$ correspond to the negative values of real part of the oscillation frequency.}\label{figworotation}
\end{figure*}

\begin{figure*}
\resizebox{\linewidth}{!}{\includegraphics*{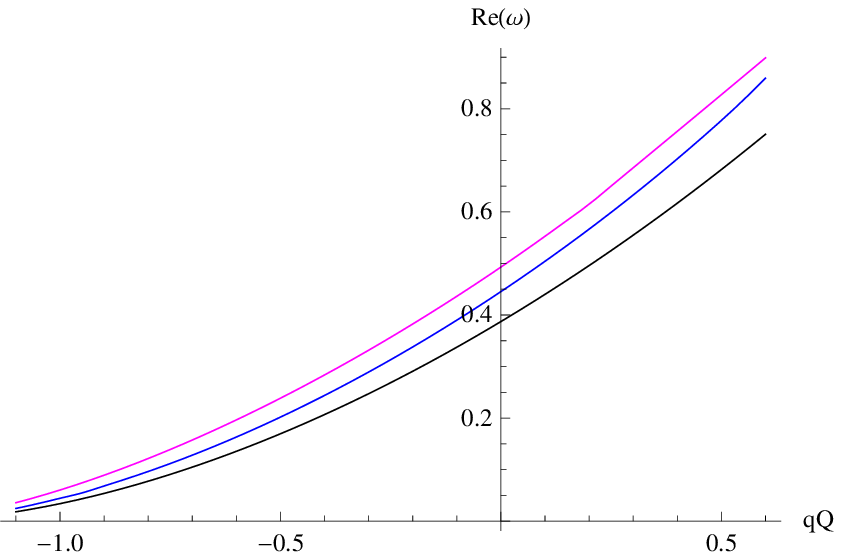}\includegraphics*{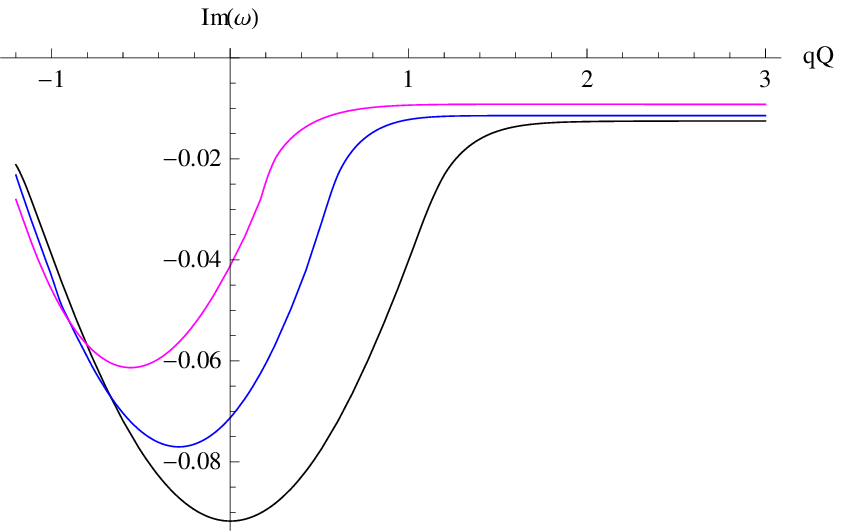}}
\caption{Real and imaginary part of the fundamental mode ($\ell=m=1$) of the massless scalar field as a function of $qQ$ for the quasi-extremal Kerr-Newman black hole ($r_-=0.95$): $a=0$ (black, bottom), $a=0.3$ (blue), $a=0.6$ (magenta, top).}\label{figrotation}
\end{figure*}

It is well known that due to a symmetry of the Kerr-Newman metric ($t\rightarrow-t$, $\phi\rightarrow-\phi$) quasinormal modes of a neutral scalar field come in within a degenerate pair
\begin{equation}\label{degeneration}
\begin{array}{rcr}
\re{\omega(m)} &=& -\re{\omega(-m)},\\
\im{\omega(m)} &=& \im{\omega(-m)}.
\end{array}
\end{equation}
When the scalar field is charged, in addition to (\ref{degeneration}) one needs to take $q\rightarrow-q$ for the symmetry to be restored. That is why we study only QNMs with positive real parts.

The degeneration (\ref{degeneration}) does not exist as soon as the coupling $qQ$ is not zero (Figs.~\ref{figworotation} and \ref{figrotation}). Then, at negative values of $qQ$, the mode with a positive real part has slower decay rate than the one with the negative one (while for positive $qQ$ the situation is opposite). Moreover, when the absolute value of $qQ$ increases (Figs.~\ref{figworotation} and \ref{figrotation}), the real oscillation frequency (given by $\re{\omega}$) approaches zero and then the mode ``disappears'' from the spectrum at $qQ$ larger than some critical value. Such disappearing of a mode in some range of parameters is not unusual and happens, for example, for arbitrarily long living modes (quasiresonances) of a massive neutral scalar field \cite{quasi-resonances}.

\subsection{Late-time tails and stability}


\begin{figure*}
\resizebox{\linewidth}{!}{\includegraphics*{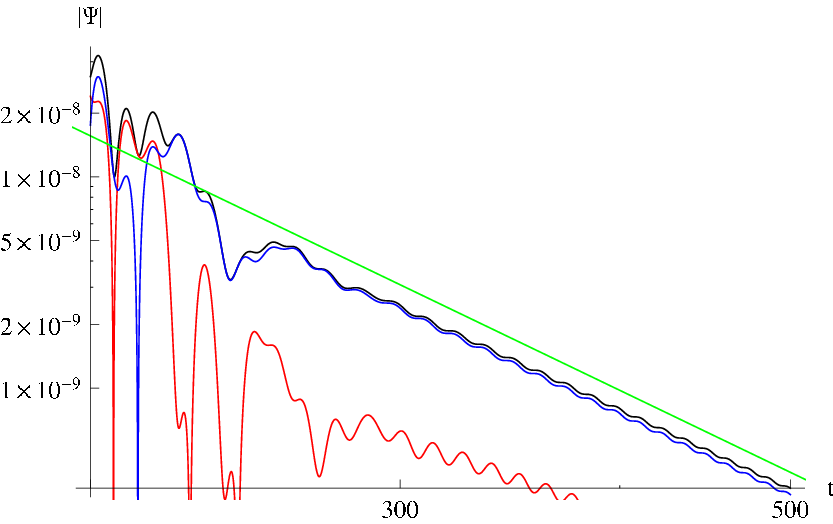}\includegraphics*{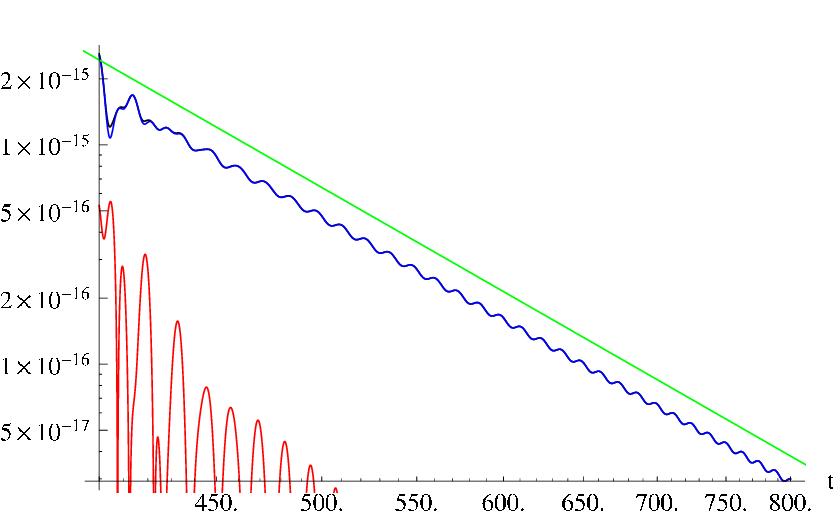}}
\caption{Late-time tails for charged ($qQ=0.5$) scalar field for $\ell=1$ (left panel) and $\ell=2$ (right panel) in the background of the extremally charged Reissner-Nordstr\"om black hole ($r_-=r_+$). Blue and red lines correspond to the real and imaginary part respectively, green lines correspond to the power-law fit $\propto t^{-2\ell-2}$.}\label{figtailsextr}
\end{figure*}


\begin{figure*}
\resizebox{\linewidth}{!}{\includegraphics*{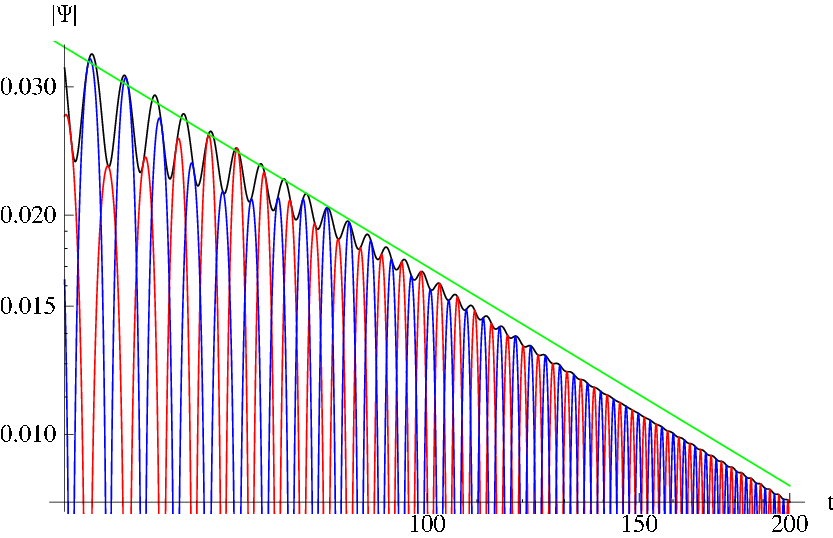}\includegraphics*{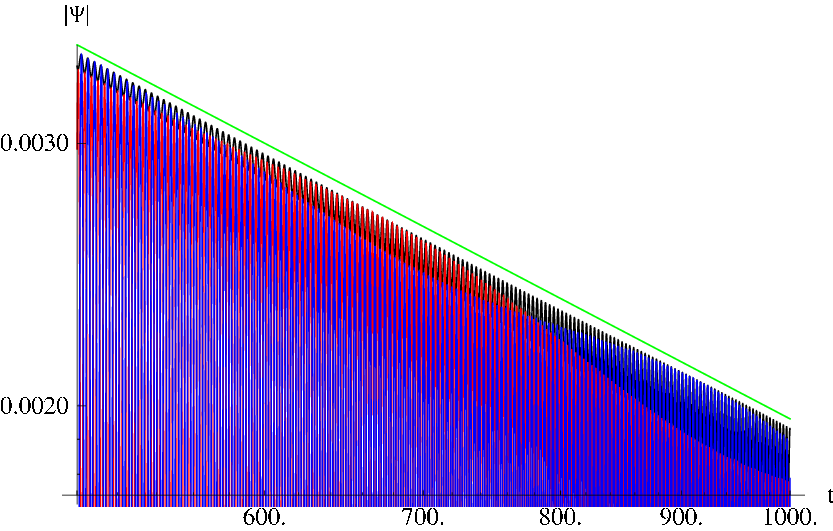}}
\caption{Asymptotic tails for
$r_-=1$,  $l=2$,  $qQ=0.5$, $\mu=1$
Blue and red lines correspond to the real and imaginary part respectively, green lines correspond to the power-law fit $\propto t^{-1}$ (left panel) and asymptotic power law $\propto t^{-5/6}$ (right panel).}\label{figasympttailsmassive}
\end{figure*}

We observe that at asymptotically late times, a neutral massless scalar field decays as
\begin{equation}
\Psi \propto  t^{-2\ell-3}, t \rightarrow \infty
\end{equation}
in concordance with \cite{Price}, while for the charged massless field, the dominant asymptotical tail is
\begin{equation}\label{chargedtail}
\Psi \propto  t^{-2\ell-2}, t \rightarrow \infty
\end{equation}
as it was first found in \cite{Bicak}. "Note, that in the regime of large $q Q$ the correction formula to (\ref{chargedtail}) was reported in \cite{Hod:1988}, which might be correct, and, then, should appear at larger values of $q Q$ than those we considered in the numerical time-domain evolution here. The late-time tails in the background of the extremally charged black hole obey the same law (see Fig.~\ref{figtailsextr}). At intermediately late times, \emph{a massive charged} scalar field decay as (see Fig.~\ref{figasympttailsmassive})
\begin{equation}
\Psi \propto  t^{-1}\sin (\mu t) ,
\end{equation}
while at asymptotically late times, the decay law is 
\begin{equation}
\Psi \propto  t^{-5/6}\sin (\mu t) , \quad t \rightarrow \infty.
\end{equation}

Thus, the power laws of a massive charged scalar field's decay at intermediate and asymptotically late times do not depend on multipole $\ell$ or charges $q$ and $Q$.  The same law was obtained in \cite{He:2006jv}  for the massive Dirac field. We suppose thereby that the asymptotic behavior of the charged massive fields does not depend on spin.

The decaying time-domain profiles for various values of the parameters $q$, $Q$ and $\mu$ show that no instability exist for nonextremal Reissner-Nordstr\"om black holes. The complementary frequency domain data gives no indication of growing modes when the rotation is not vanishing.
Therefore, we conclude that there are no signs of instability for a massive charged scalar field in the nonextremal Kerr-Newman background under the quasinormal modes' boundary conditions.

\section{QNMs of a neutral field for nearly and exactly extremal black holes: mode branching and stability}

\subsection{A massless scalar field}

We shall measure all quantities in units of $r_{+}$, that is we take $r_+=1$. In order to keep staying near the extremal state, it is sufficient to keep $r_{-}$ close to $r_{+}$.

Here we shall start from the generalization of level plots of \cite{Yang:2012pj}, which show the branching of modes.
In \cite{Hod:2007} it was found that, in the near-extremal Kerr limit, Zero-Damped Mode (ZDM) satisfies
\begin{equation}\label{eqhod2}
\omega = \frac{m}{2}-\frac{\delta \sqrt{\epsilon}}{\sqrt{2}}-i\left (n+\frac{1}{2}\right ) \frac{\sqrt{\epsilon}}{\sqrt{2}} +{\cal O}(\epsilon), \qquad \epsilon\rightarrow0,
\end{equation}
where $\epsilon=1-a$, $\delta=\sqrt{(7m^2-1)/4-\lambda(\omega=m/2)}$.

\begin{figure*}
\resizebox{\linewidth}{!}{\includegraphics*{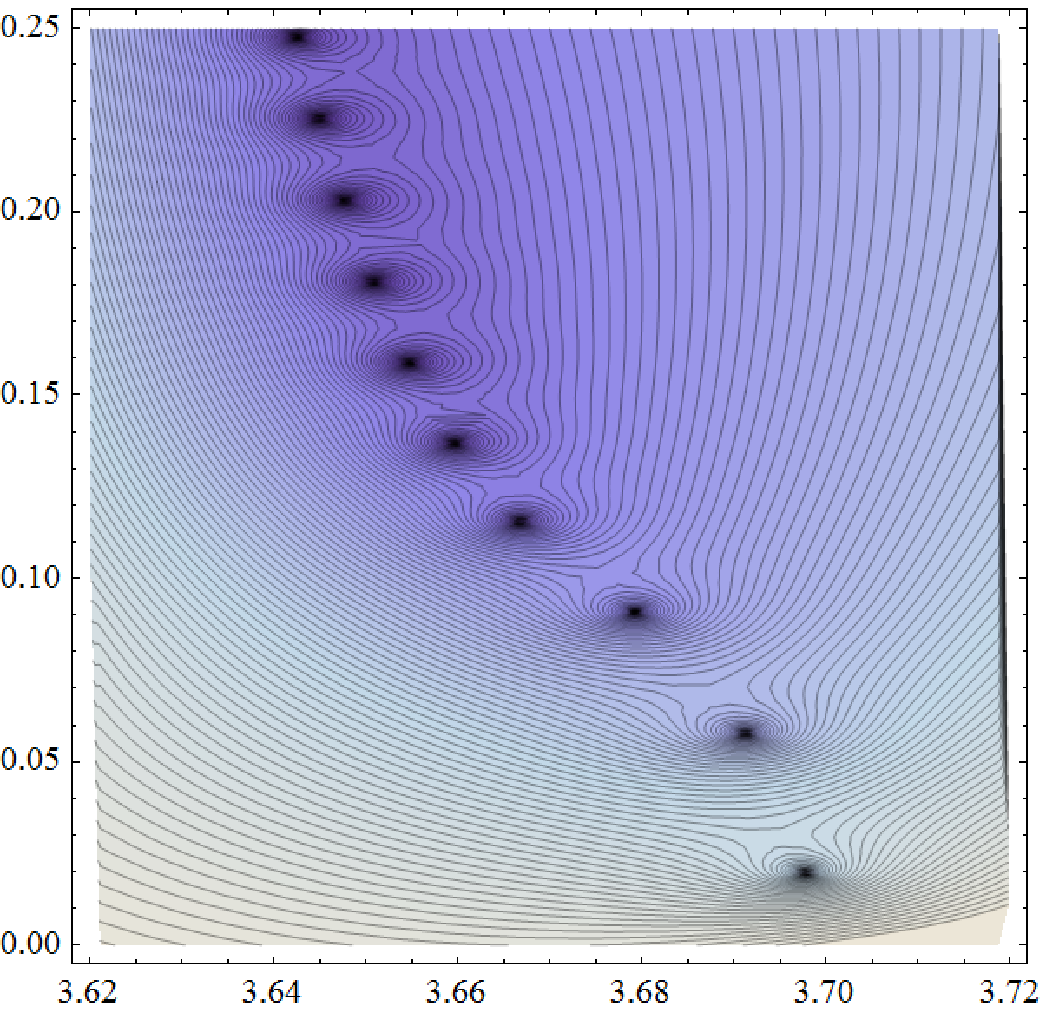}\includegraphics*{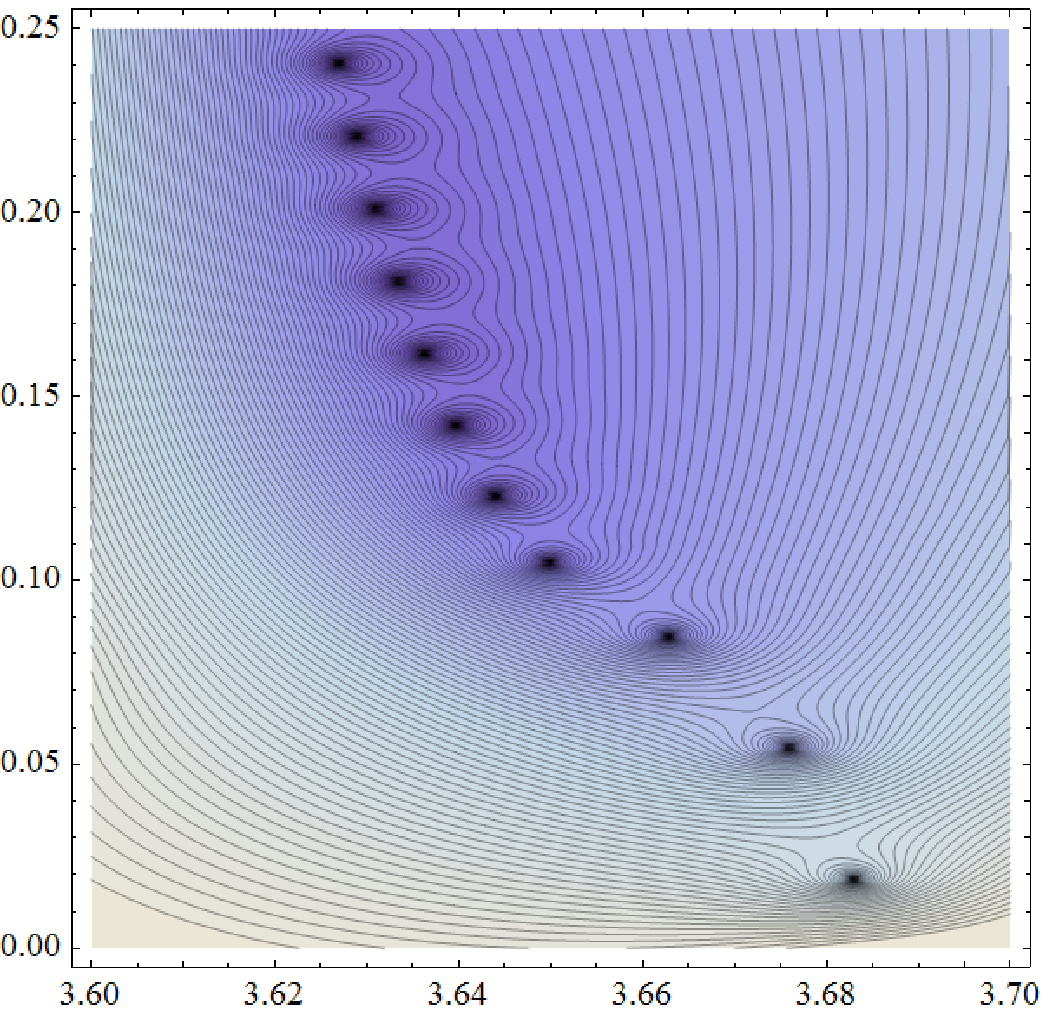}\includegraphics*{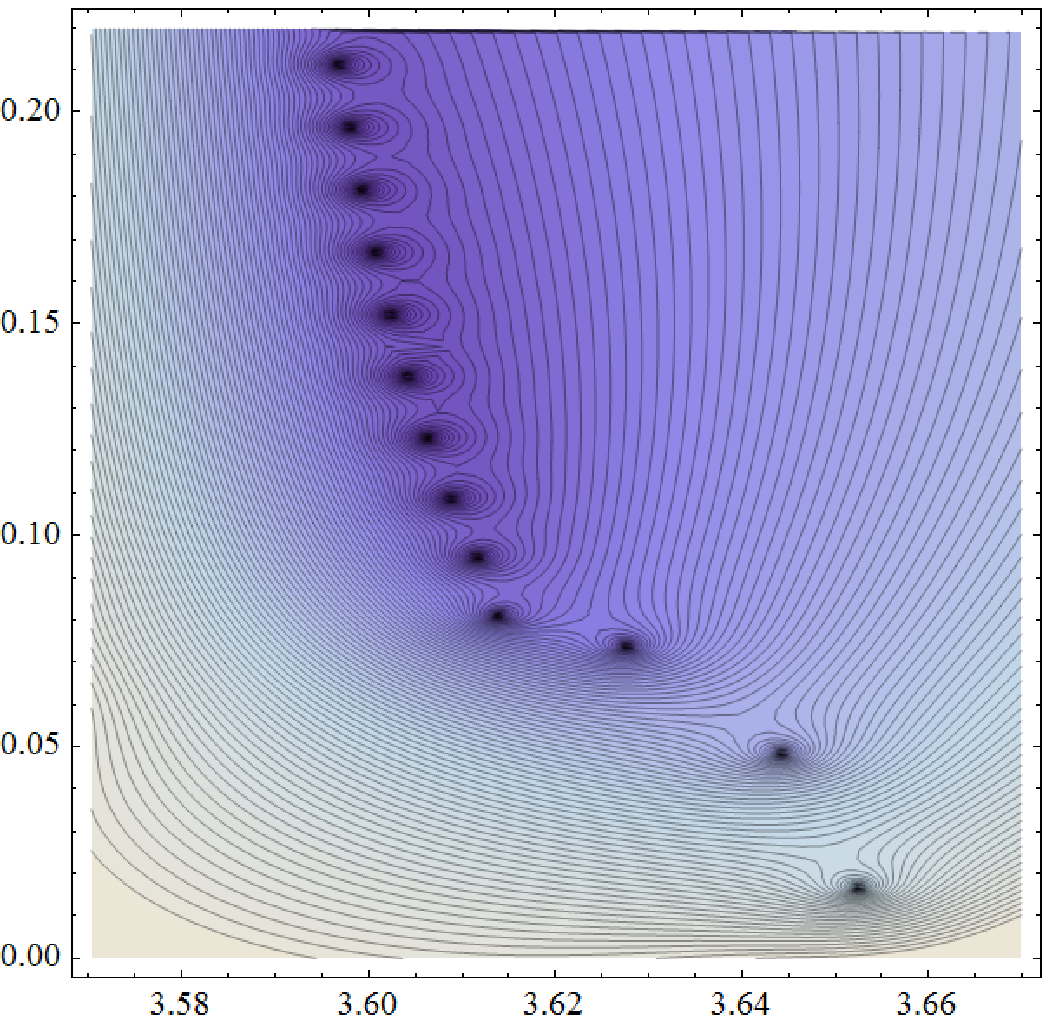}}
\resizebox{\linewidth}{!}{\includegraphics*{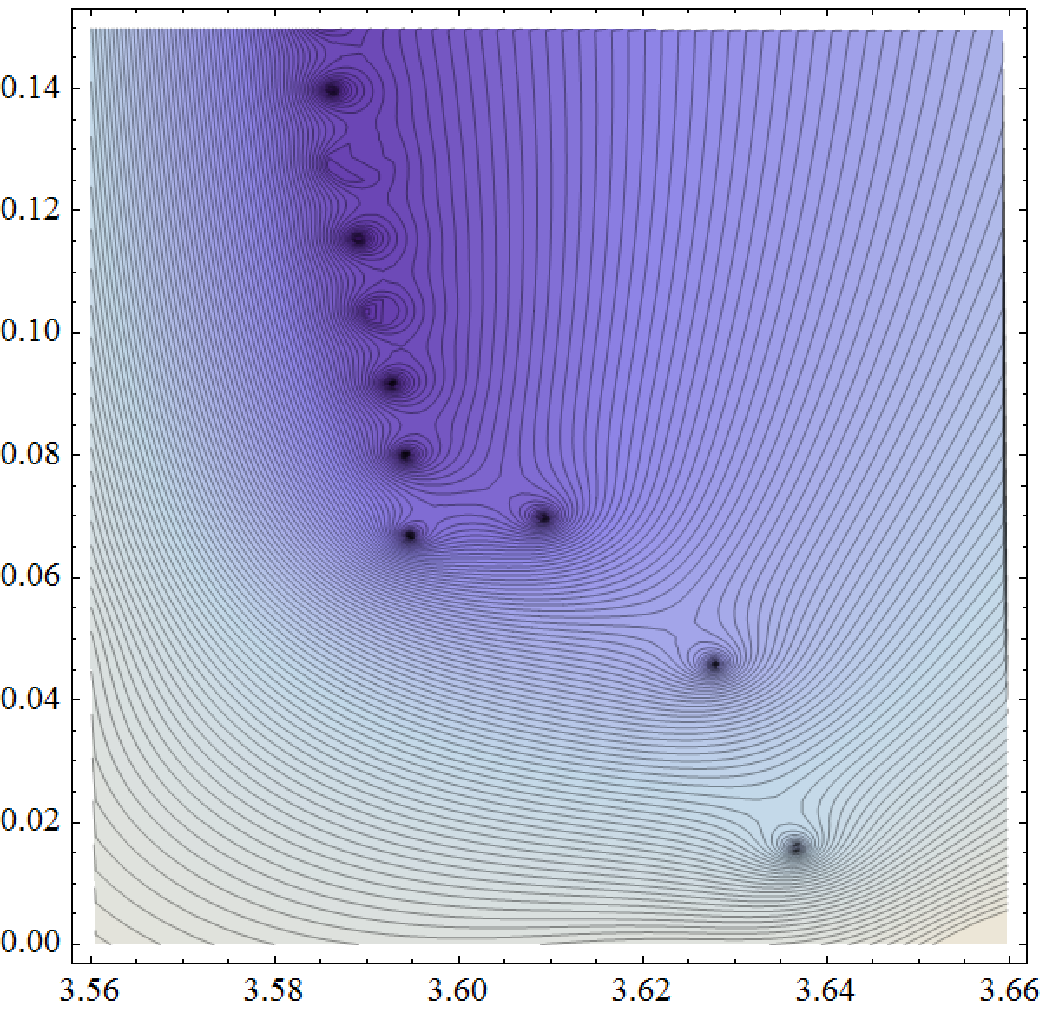}\includegraphics*{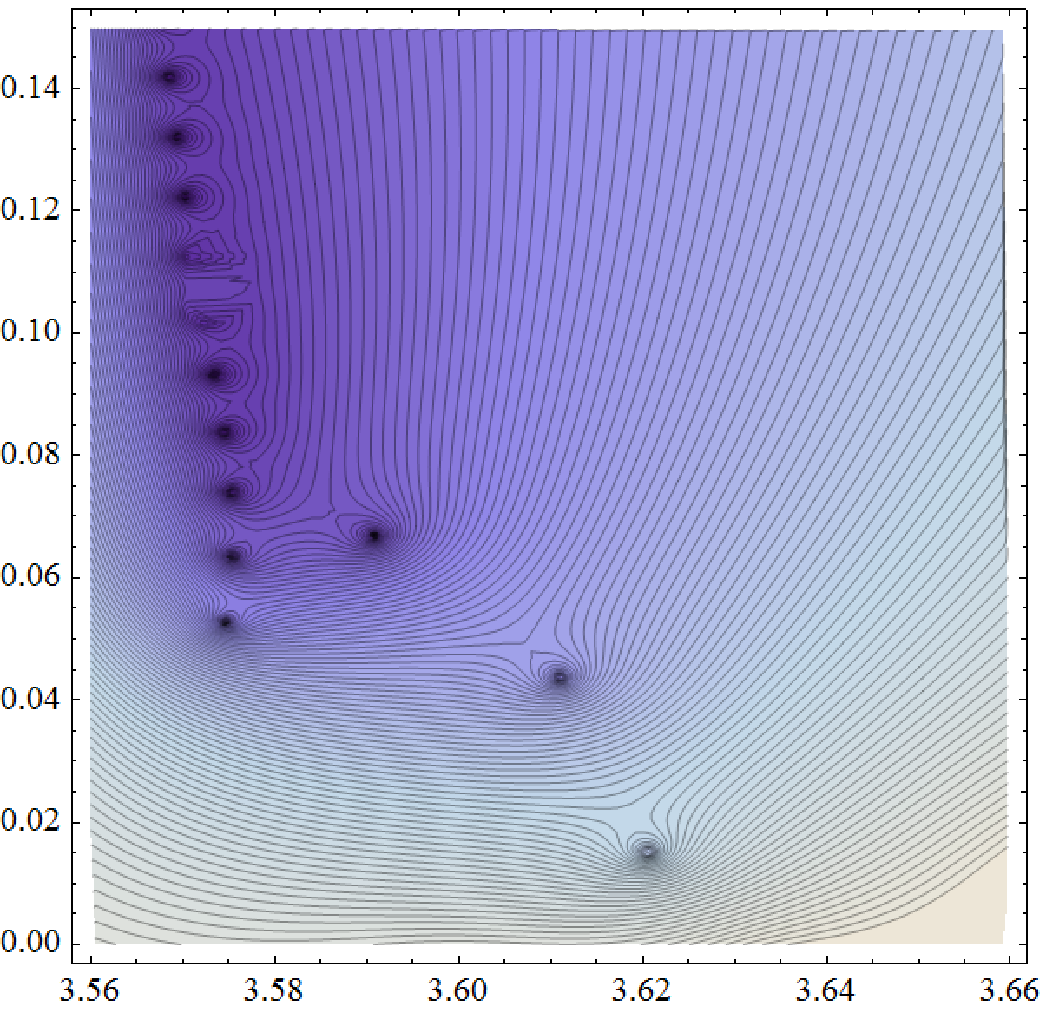}\includegraphics*{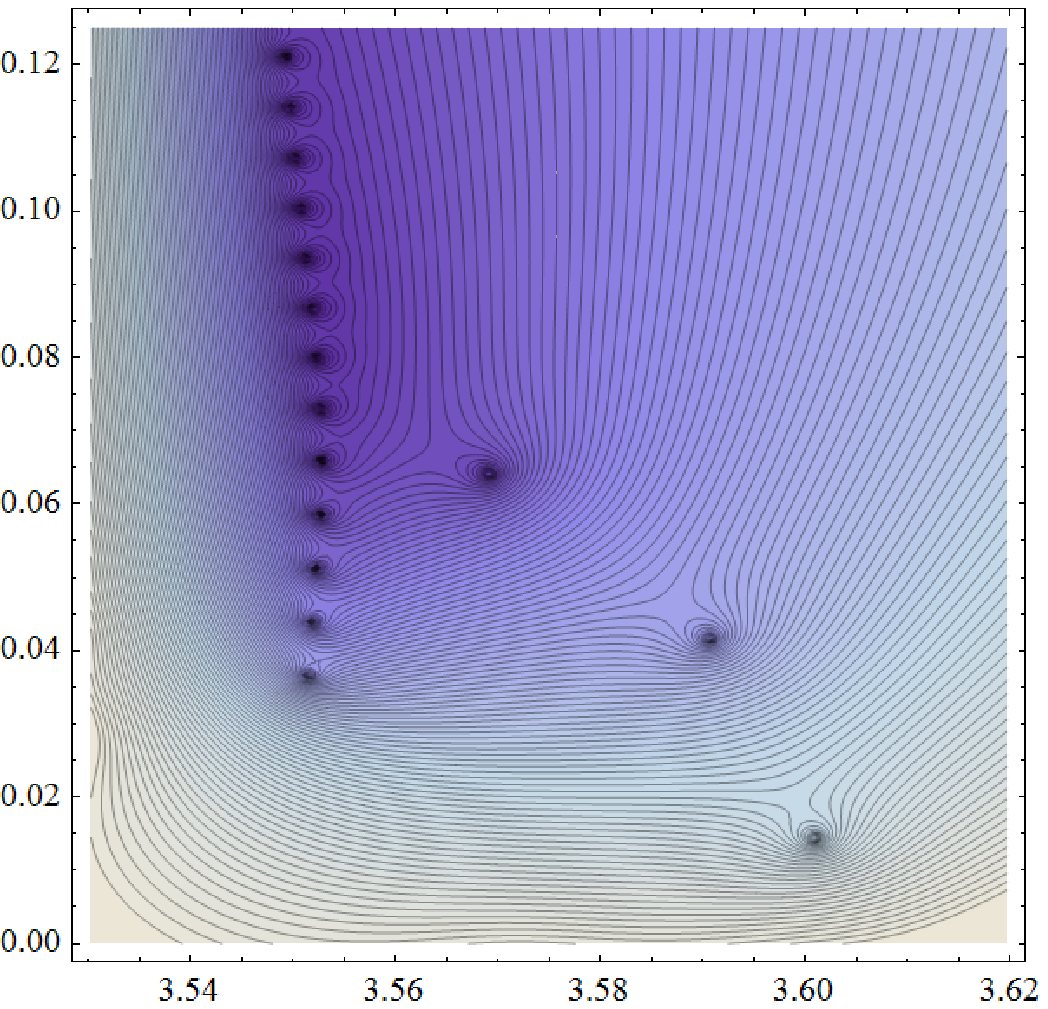}}
\caption{Level plots of the logarithm of the absolute value of the difference between the left and right sides of the equation with continued fraction as a function of the real (horizontal axis) and imaginary (vertical axis) parts of the frequency for the scalar field ($\ell=10$, $m=7$) in the background of a near-extremal Kerr-Newman black hole ($r_{-}=0.972111205849$). From left to right: $a=0.955$, $a=0.960$, $a=0.970$, $a=0.975$, $a=0.980$, $a=0.985957$ (extremal rotation).}\label{fig1}
\end{figure*}

Therefore, in Fig.~\ref{fig1} the same multipole $\ell$ and azimuthal $m$ numbers were  chosen as in  \cite{Yang:2012pj}. In Fig.~\ref{fig1} it is shown the logarithm of the absolute value of  difference between the left and right hand sides of the continued fraction equation (\ref{continued_fraction}) as a function of real and imaginary parts of $\omega$.
In Fig.~\ref{fig1} we can see that at some near-extremal $r_{-}$, and $a$, which starts from $a=0.955$ until its extremal value, the new branch of modes appears: the two branches of modes, which take place for high rotation ($a \gtrapprox 0.970$), merge into a single one at slower rotation ($a \lessapprox 0.960$).  As it is expected, for small values of black hole's charge $Q$, the obtained in Fig.~\ref{fig1} Zero-Damped Mode is close to the one described by Eq.~(\ref{eqhod2}). The mode branching is owing to extremal rotation and not owing to extremal charge. Indeed, a neutral scalar field around the near-extremal Reissner-Nordstr\"om black hole shows no mode branching.

We have checked that the time-domain profile of scalar and gravitational perturbations of the exactly extremal Reissner-Nordstr\"om black hole
consists of damped quasinormal oscillations, what proves stability of the extremal Reissner-Nordstr\"om solution from the point of view of an external observer, because 1) the profile includes contribution from all the modes, so that none can be missed, 2) the method of calculations does not imply any specific boundary conditions at the event horizon. This does not contradict a special instability of the extremal Reissner-Nordstr\"om black holes \cite{Aretakis:2011ha,Aretakis:2011gz} as this instability develops only on the event horizon and cannot be observed by an external observer \cite{Lucietti:2012xr}.

\begin{table}
\caption{Fundamental quasinormal modes ($\ell=0$) of the scalar field for the extremal Kerr-Newman black hole ($r_+=r_-=1$).}\label{tablI}
\begin{center}
\begin{tabular}{|c|c|}
\hline
$a$&$\omega$\\
\hline
$0.0$&$0.133459 - 0.095844\imo$\\
$0.1$&$0.133124 - 0.095779\imo$\\
$0.2$&$0.132137 - 0.095582\imo$\\
$0.3$&$0.130555 - 0.095252\imo$\\
$0.4$&$0.128456 - 0.094788\imo$\\
$0.5$&$0.125938 - 0.094188\imo$\\
$0.6$&$0.123100 - 0.093458\imo$\\
$0.7$&$0.120035 - 0.092605\imo$\\
$0.8$&$0.116826 - 0.091641\imo$\\
$0.9$&$0.113554 - 0.090578\imo$\\
$1.0$&$0.110246 - 0.089433\imo$\\
\hline
\end{tabular}
\end{center}
\end{table}


The exactly extremal Kerr-Newman black hole is also stable against scalar-field perturbations, as it can be seen from the damped fundamental quasinormal modes on Table~\ref{tablI}. The real oscillation frequency and the damping rate of the fundamental $\ell=m=n=0$ mode monotonically decrease, as the rotation parameter $a$ grows.
The quasinormal modes of  the near-extremal Kerr-Newman black hole approach their exactly extremal values.  The $\re{\omega}$  and $\im{\omega}$ monotonically decreases, as $r_{-}$ grows in the near-extremal regime.

\subsection{A massive scalar field}

A massive field has a number of qualitative distinctions from the massless case. First, there are arbitrarily long living modes -- quasiresonances --
which approach asymptotically the bound states at the zero damping rate limit. Let us explain this in more detail. ``Ordinary'' quasinormal modes with however small but finite damping rate satisfy the boundary condition of purely ingoing wave at the horizon and at infinity. This means that the wave function diverges at the horizon. Yet, when $\im{\omega}$ approaches zero, energy conservation requires that the amplitude vanishes both at the event horizon and at infinity, which means approaching the \emph{bound-state} problem in the limit $\im{\omega}=0$. As quasiresonances have arbitrarily small but nonzero damping rate, they are not bound states. However, their relation to another phenomena, \emph{superradiance} \cite{Starobinsky}, could be essential. A superradiance is a channel of extraction of rotational energy from a black hole by a wave reflected from it with larger amplitude than the initially incident wave. When the field has mass $\mu$,  a local minimum of the effective potential appears far from the black hole. The repeated reflection from the local maximum could create an instability, yet, as it was shown in \cite{Konoplya:2006br} for the Kerr solution, a massive scalar field is stable under the quasinormal modes' boundary conditions, so that, in order to gain an instability, a Dirichlet boundary condition (that is an effective confining box for perturbation) far from the black hole must be imposed. Such a confining condition is asymptotically achieved in quasiresonances, so that, if quasiresonant modes were superradiant, an instability could be expected. Apparently, in the near-extremal regime, the mode branching would ``interfere'' with quasiresonances. Therefore, we shall consider here a massive scalar field for the near-extremal rotation in detail.

\begin{figure*}
\resizebox{\linewidth}{!}{\includegraphics*{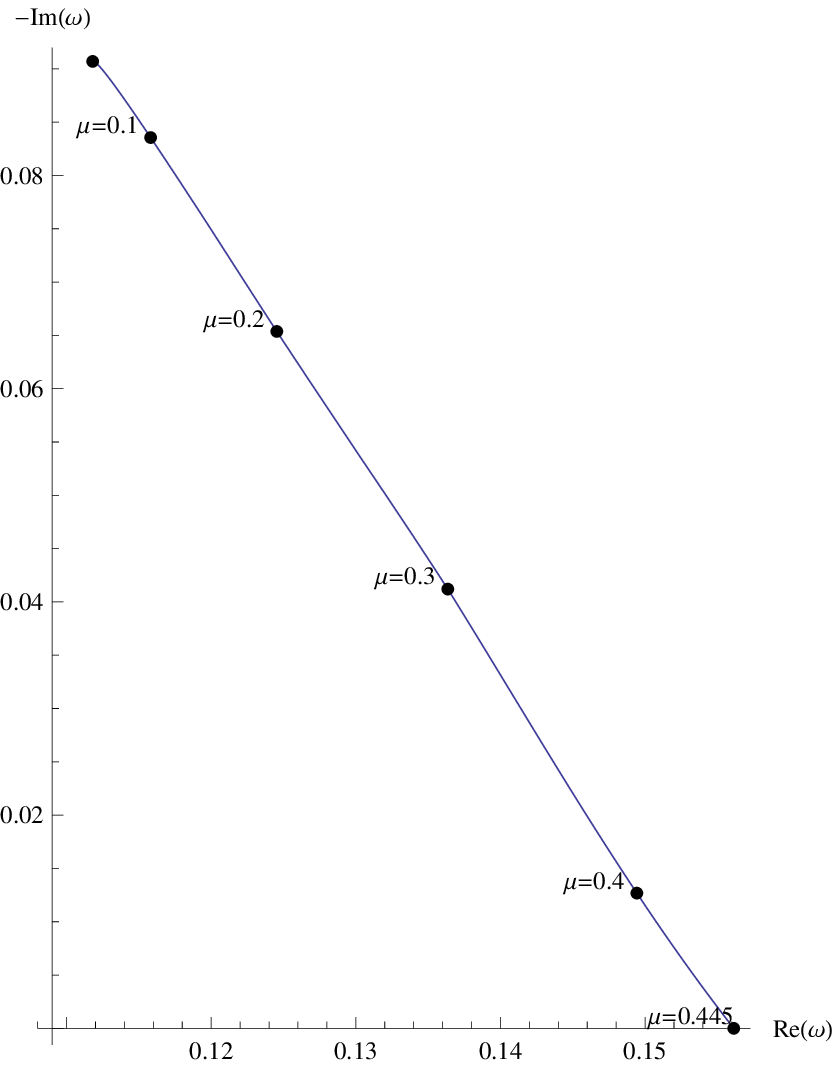}\includegraphics*{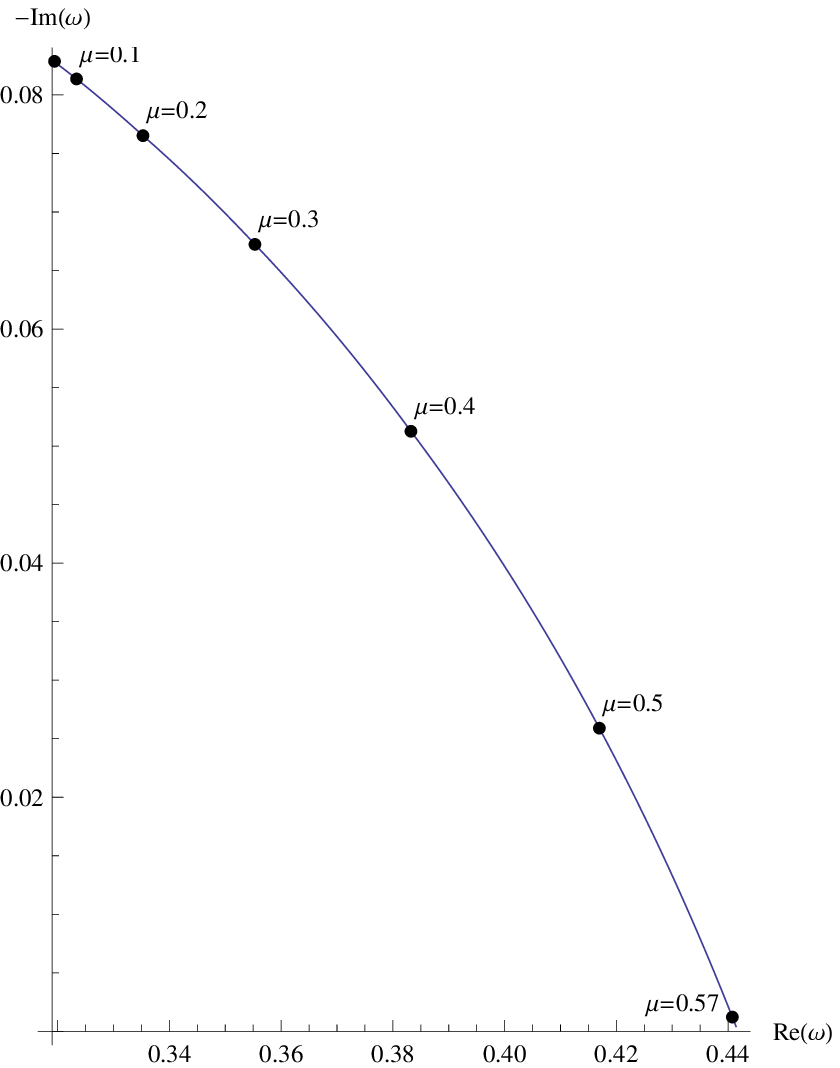}\includegraphics*{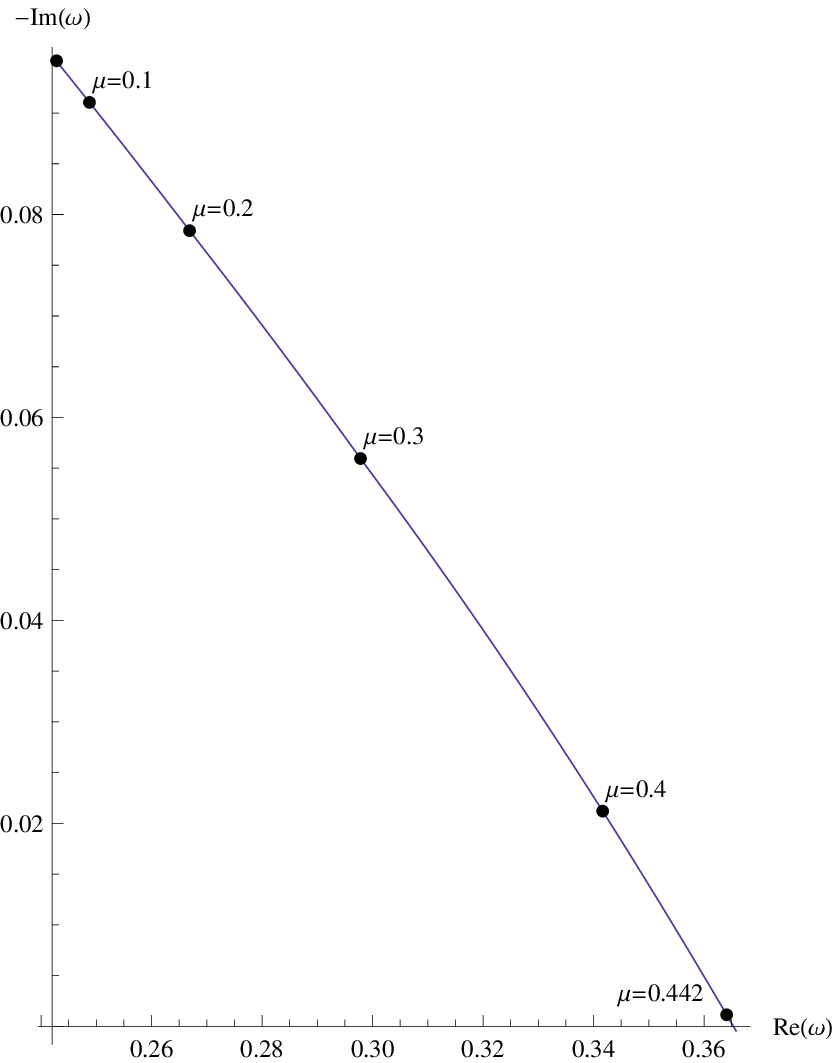}}
\caption{The fundamental mode as a function of mass for a near-extremal Kerr black hole ($r_{-}=0.972111205849$, $a=0.985957$): $\ell=m=0$ (left panel); $\ell=1$, $m=0$ (middle panel); $\ell=1$, $m=-1$ (right panel). }\label{fig5}
\end{figure*}

\begin{figure*}
\resizebox{\linewidth}{!}{\includegraphics*{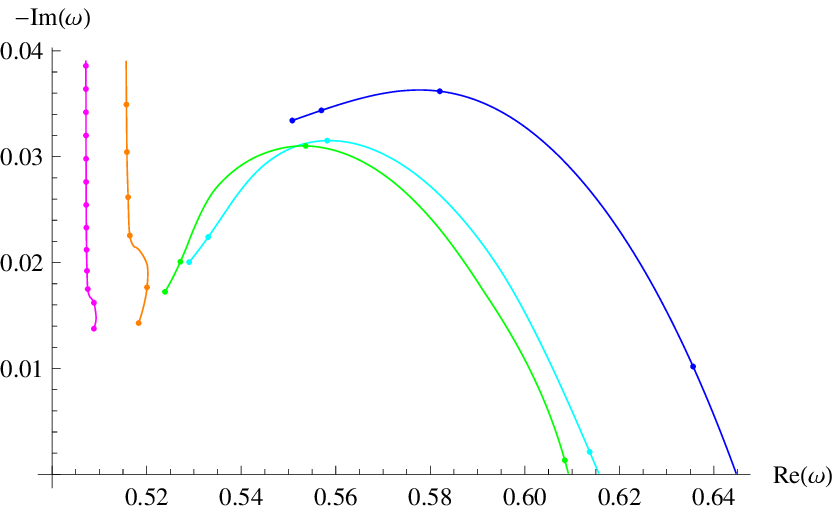}\includegraphics*{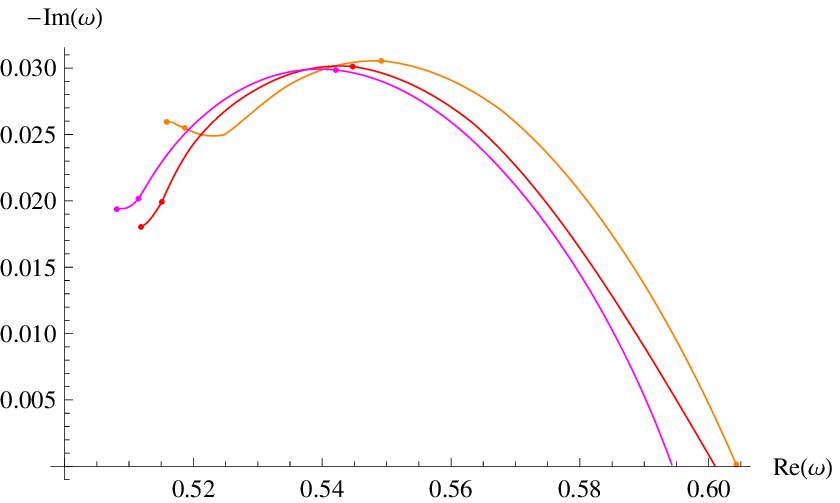}}
\caption{Left plot: Quasinormal modes which are fundamental at $\mu =0$, parameterized by mass (dotted values: $\mu=0$, $\mu=0.3$, $\mu=0.6$, $\mu=0.9$...) from right to left: a=0.90 (blue), a=0.95 (cyan), a=0.96 (green), a=0.97 (orange), a=0.985957 (magenta). Right plot: Quasinormal modes which are first overtones at $\mu =0$ from right to left: a=0.97 (orange), a=0.98 (red), a=0.985957 (magenta). The orange and magenta lines show that for the selected values of rotation and mass the initially (at $\mu =0$) fundamental mode becomes the first overtone. The quasiresonances exist for all values of $\mu$ and $a$. }\label{fig6}
\end{figure*}

As it is illustrated in Fig.~\ref{fig5}, quasiresonances exist even for very quick and near-extremal rotation for nonpositive values of $m$. For positive $m$ and near-extremal rotation, we observe that as the field mass $\mu$ grows, the fundamental mode (with larger oscillation frequency and the slowest damping rate) does not go over to the quasiresonance limit while the first overtone does (see Fig.~\ref{fig6}). Therefore, at some value of the field mass $\mu$ the first overtone has the same decay rate as the fundamental mode. For masses above this threshold the fundamental mode and  first overtone ``exchange'': the lower-damped mode corresponds to the first overtone of the massless field. Thus, the actual oscillation frequency of the
lowest damping mode as a function of the field mass has a discontinuity at this point.

An extensive search in the frequency domain has not shown any signs of instability of a neutral massive scalar field in the near-extremal regime up to masses $\mu M \sim 3$ .

\section{QNMs of a charged field: technical difficulties in the near extremal and higher overtone regimes}

\begin{figure*}
\resizebox{\linewidth}{!}{\includegraphics*{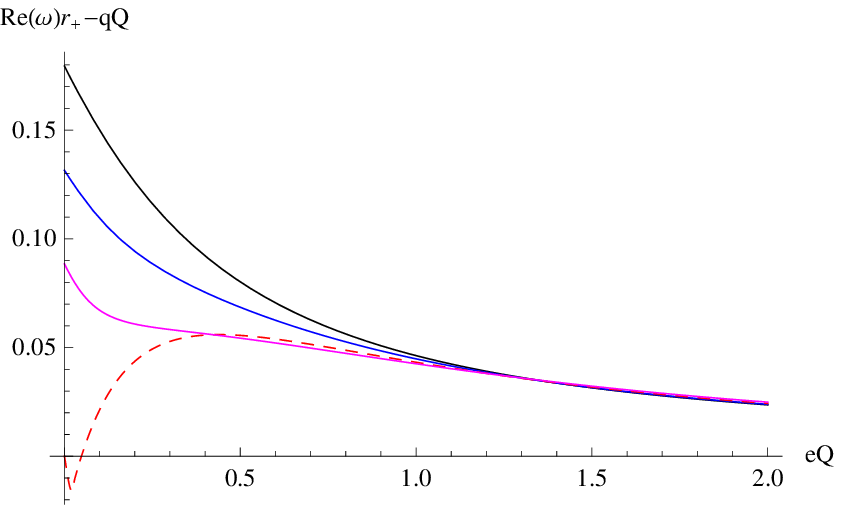}\includegraphics*{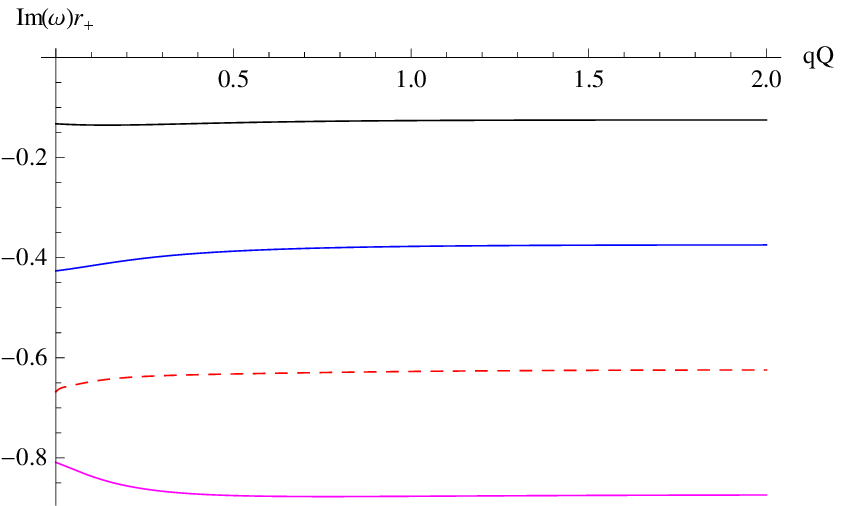}}
\resizebox{\linewidth}{!}{\includegraphics*{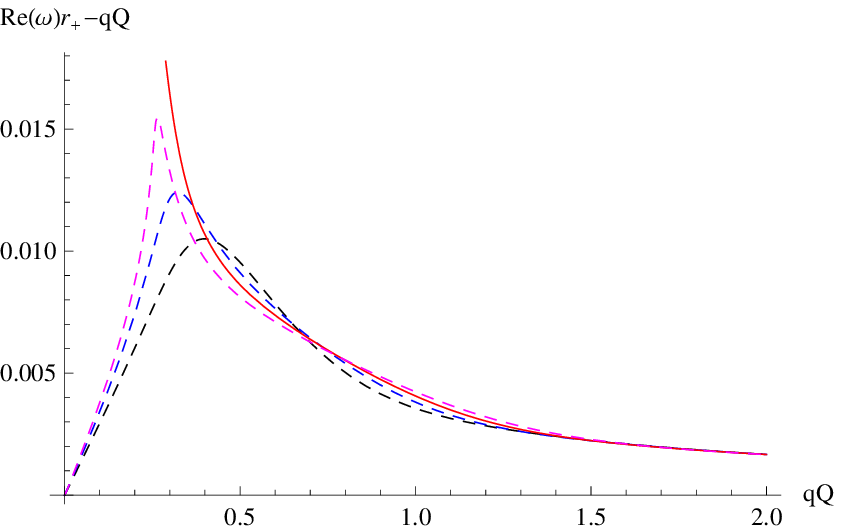}\includegraphics*{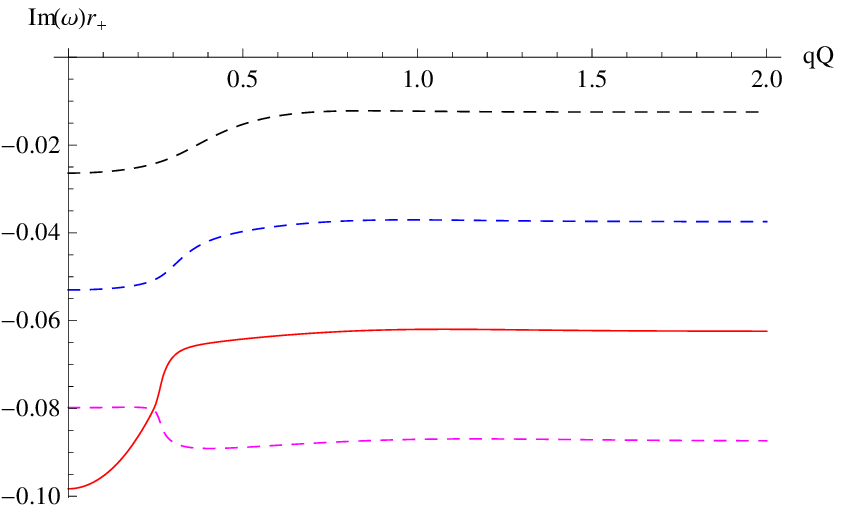}}
\caption{Real (left panels) and imaginary (right panels) part of the four dominant QNMs of the charged scalar field ($\ell=0$) for the Reissner-Nordstr\"om black hole: $r_-=0.50r_+$ (top) and $r_-=0.95r_-$ (bottom). Dashed lines mark the modes which correspond to the purely imaginary limits for the noncharged scalar field, that are not solutions to (\ref{continued_fraction}). For large field charge $q$ the frequencies satisfy the analytical formula (\ref{asymptotical_frequency}).}\label{figimaginarymode}
\end{figure*}

When the value of the black hole's charge is not very close to its extremal values, say, up to around $Q \approx 0.99M$, we were able to compute the dominant quasinormal frequencies for the full range of parameters $Q$, $q$, $a$, $\mu$, and, in the limit of vanishing rotation, QN frequencies extracted from the time-domain profiles with the help of Prony method were in a good agreement with the modes obtained by the Frobenius method.
Yet, when the field's charge $q$ is not zero, one meets  two difficulties when he is trying to compute higher overtones or to approach quite closely the extremal regime. The first problem is the existence of new poles of the continued-fraction equation (\ref{continued_fraction}) (shown as dashed lines in Figs.~\ref{figimaginarymode}) for nonzero $q$ which have odd behavior: 
$\re{\omega}$ of these new ``modes'' proportional  approximately to $qQ/r_+$ for small and moderate $q Q$. In the limit $q Q \rightarrow 0$, $\re{\omega}$ does not go zero, because the ``mode''  disappears at some small critical value of $qQ$. For larger values of $Q$, the critical value of $qQ$ is smaller. Thus, for $r_{-} =0.5$, $(qQ)_{crit} \approx 0.18$, while for $r_{-} =0.95$, $(qQ)_{crit} < 0.01$.
As these special poles of Eq.~(\ref{continued_fraction}) do not go over into quasinormal modes of a neutral scalar field when $q \rightarrow 0$, they should be checked by alternative calculation. These strange ``modes'' exist for all values of the black-hole charge $Q$, yet, at smaller $Q$ they correspond to higher overtones which therefore decay very quickly.

\begin{table}
\caption{The four dominant QNMs for the Reissner-Nordstr\"om black hole ($r_-=0.5$).}
\begin{tabular}{|c|rl|r|}
\hline
mode&\multicolumn{2}{|c|}{time-domain fit}&\multicolumn{1}{|c|}{Leaver}\\
\hline
$n=0$&$0.17937$&$-0.13257\imo$&$0.17945-0.13250\imo$\\
$n=1$&$0.122$&$-0.430\imo$&$0.13151-0.42639\imo$\\
im. limit&&?&$-0.66796\imo$\\
$n=3$&$0.1$&$-0.8\imo$&$0.08846-0.80911\imo$\\
\hline
\end{tabular}\label{problems}
\end{table}

The second problem, already mentioned above, is absence of convergence of the Prony method in the regime when modes with very small real part dominate in the spectrum. This makes it impossible to check with the time-domain integration if the poles observed through the Frobenius method are true quasinormal modes or not (see table~\ref{problems}). Nevertheless, this technical problem does not affect the issue of stability, as neither frequency domain nor time domain show any signs of growing modes.

\begin{figure*}
\resizebox{\linewidth}{!}{\includegraphics*{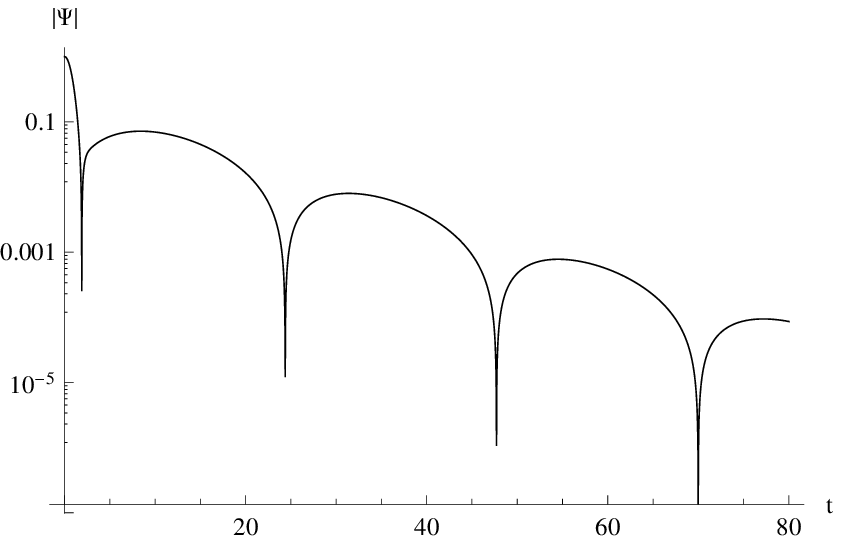}\includegraphics*{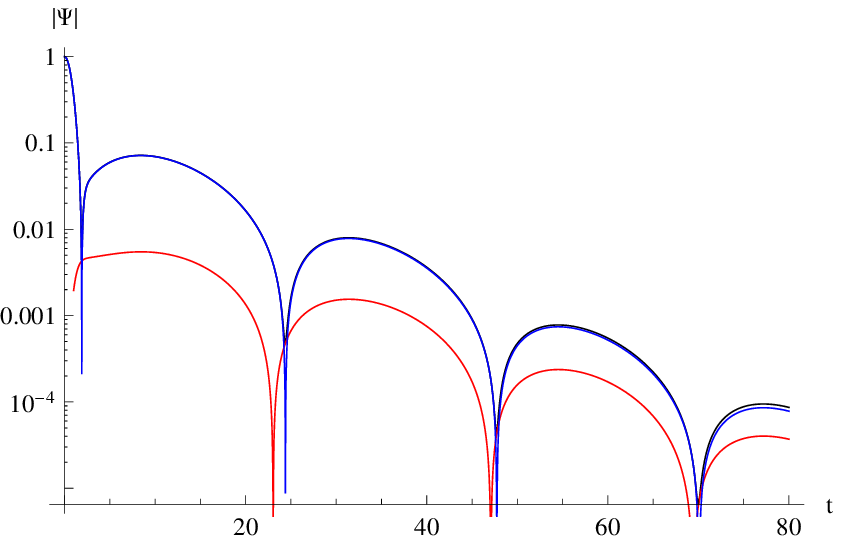}}
\caption{The time domain profiles of the neutral scalar field field (left panel) and for a charged ($qQ=0.01$) scalar field (right panel) for the Reissner-Nordstr\"om black hole ($r_-=0.95$). On the right panel blue and red lines correspond to the real and imaginary part respectively. For the neutral field the time-domain fit allows to find the dominant frequency $\omega=0.13697 - 0.09832\imo$, which is close to the accurate result found by the Leaver method $\omega=0.13688-0.09830\imo$. For the charged field from the time-domain fit we find that the dominant frequency with the positive real part is $\omega=0.14215-0.09829\imo$ which corresponds to the third overtone found with the help of the Leaver method $\omega_3=0.14203-0.09827\imo$, while the three dominant frequencies ($\omega_0=0.01029-0.02639\imo$, $\omega_1=0.01032-0.05299\imo$, $\omega_2=0.01032-0.07980\imo$) are not observed. The real parts of these frequencies are close to $qQ$, approaching purely the imaginary limit for the noncharged field.}\label{figtimedomainproblem}
\end{figure*}

In order to find the dominant modes accurately with the help of the Prony method, one must fit the time-domain profile by the exponents (\ref{damped-exponents}) in the certain time interval when the quasinormal ringing is observed. The interval must be chosen in such a way that the initial outburst is already damped and, at the same time, the late-time tail is still not dominant in the signal. The natural criterium to learn if we chose the correct time interval, is convergence of indexes of the exponents (obtained modes) with respect to the increasing number of exponents $p$ in (\ref{damped-exponents}). At the quasinormal-ringing stage the dominant frequencies do not depend on $p$ because contributions of higher overtones are exponentially damped. Therefore, we can fit the time-domain profiles in such time intervals which provide convergence with respect to $p$. However, for a charged field (or even for a neutral one if the black hole is highly charged (as in Fig.~\ref{figtimedomainproblem})), in addition to the convergent complex frequencies, which correspond to the quasinormal modes, we find damped purely imaginary frequencies, which do not converge with respect to $p$, but nevertheless have large amplitude. Appearing of these frequencies in the fit could be an indication of existence of a number of (almost) purely imaginary modes in the quasinormal spectrum which cannot be well fit by only a few exponents. Yet, these ``modes'' could appear due to a remnant noise from the initial stage or due to the subdominant contributions from the late-time tails that might reveal themselves at the quasinormal stage. We believe that a further thorough study of this question could tell if these frequencies are a new kind of quasinormal modes with vanishing real part or just a numerical artifact.

\section{Conclusions}

In this paper we have generalized results of Yang et al. \cite{Yang:2012pj} on mode branching of Kerr space-time in two ways. We considered quasinormal modes of a neutral massive scalar field around the near-extremal Kerr-Newman black hole and showed that a similar branching takes place for the near extremal rotation. The fundamental quasinormal modes of the Kerr-Newman black hole are shown to be damped which implies stability and quasinormal modes of near extremal black hole approach their extremal values. We also showed that arbitrarily long living modes, quasiresonances, exist, obey mode branching and show no sign of instability.

These results do not contradict the instability found in \cite{Aretakis:2011ha,Aretakis:2011gz}, because the quasinormal mode formulation of the perturbation problem implies that the perturbation is formed and propagates outside the event horizon, while the instability found in \cite{Aretakis:2011ha,Aretakis:2011gz} propagates only along the null cone.
Thus, an external observer will never see such an instability, which, apparently, could be seen by an in-falling into the black hole observer.
The linear stability of the extremal configuration for an external observer is illustrated here by damped time-domain profiles of perturbation propagating outside the extremal Reissner-Nordstr\"om black hole (see Fig.~\ref{figtailsextr}). However, nonlinear perturbations might produce an instability which develops in the time cone as well, leading to a physical consequence for an external observer \cite{Murata-com}.

Our main conclusion here is that a massive charged scalar field in the Kerr-Newman background is stable for any physically meaningful values of the parameters $q$, $\mu$, $Q$, $a$, $M$, because an extensive search of quasinormal modes showed no signs of growing modes in the spectrum.
We have found that at some values of $qQ$, the fundamental quasinormal mode may have vanishing real part. This does not happen for neutral fields.

\begin{table}
\caption{Summary of power-law decays of a scalar field in Schwarzschild and Reissner-Nordstr\"om backgrounds.}
\label{tail-summary}
\begin{tabular}{|c|c|c|}
  \hline
  $q, \mu$ & $Q=0$ & $Q \leq M$ \\
  \hline
  $q=0, \mu=0$ & $t^{- 2 \ell -3}$ & $t^{- 2 \ell -3}$ \\
    \hline
  $q=0, \mu \neq 0$ & $t^{-5/6} \sin(\mu t)$ & $t^{-5/6} \sin(\mu t)$ \\
    \hline
  $q \neq 0, \mu=0$ & -- & $t^{- 2 \ell -2}$ \\
    \hline
  $q \neq 0, \mu \neq 0$ & -- & $t^{-5/6} \sin(\mu t)$ \\
   \hline
\end{tabular}
\end{table}

We have also showed that the decay of a scalar field (be it charged or neutral, massive or massless) is dominated by the power-law tails (see table (\ref{tail-summary})) which are the same for the nonextremal black hole as for the extremal one. Here we have considered an already formed black hole and the perturbation propagating outside of it, while the modeling of collapse of charged matter leads to different result for the late-time decay of the extremal Reissner-Nordstr\"om black hole \cite{Bicak}. Thus we complement earlier results on late-time tails of neutral massive \cite{Koyama:2001ee} and  massless fields \cite{Bicak} in the Schwarzschild and Reissner-Nordstr\"om backgrounds.

In addition, when $q \neq 0$ the continued fraction equation (\ref{continued_fraction}) indicates presence of some new roots in the frequency domain, which do not exist for the neutral field, and whose real parts are proportional to $qQ/r_+$. These roots exist for any value of $Q$ and $a$ (and nonzero $q$), but correspond to higher ``overtones'' for weakly charged black hole (Fig.~\ref{figimaginarymode}), yet, they become dominant for nearly extremal black holes. These ``modes'', if they exist, cannot be extracted from a time-domain profile, probably due to the quick onset of asymptotic tails, and, therefore, must be further verified with an alternative method of calculation.

The obtained here conclusion on the stability of a massive charged scalar field around Kerr-Newman black hole allows us to go on the study of Hawking radiation for this case. Using numerical techniques makes it possible to find grey-body factors accurately for the full range of parameters \cite{GKZ2}.

\section*{Acknowledgments}
This work was supported by the European Commission grant through the Marie Curie International Incoming Program.
R. A. K. acknowledges support of his visit to Universidade Federal do ABC by FAPESP.
A.~Z. was supported by Conselho Nacional de Desenvolvimento Cient\'ifico e Tecnol\'ogico (CNPq).

\end{document}